\documentclass[fleqn,usenatbib]{rasti}

\usepackage{newtxtext,newtxmath}
\usepackage[T1]{fontenc}

\DeclareRobustCommand{\VAN}[3]{#2}
\let\VANthebibliography\thebibliography
\def\thebibliography{\DeclareRobustCommand{\VAN}[3]{##3}\VANthebibliography}

\usepackage{graphicx}	% Including figure files
\usepackage{amsmath}	% Advanced maths commands
\usepackage{xspace}
\usepackage{threeparttable}
\usepackage{orcidlink}
\usepackage{xcolor}

\newcommand{\sgr}{SGR~1935+2154\xspace}
\newcommand{\leuschsize}{4.3-m\xspace}
\title[LIMBO]{Long-Integration Magnetar Burst Observatory (LIMBO): Instrument Summary and Early FRB Rate Constraints}

\author[D. McCauley et al.]{
Darby McCauley \orcidlink{0009-0003-2691-4222},$^{1,2}$\thanks{E-mail: darbynm2@illinois.edu}
Aaron Parsons \orcidlink{0000-0002-5400-8097},$^{2,3}$
Wei Liu \orcidlink{0000-0002-9131-3664},$^{3}$
Wenbin Lu \orcidlink{0000-0002-1568-7461},$^{2,4}$
Dirk Wright,$^{3}$
and Dan Werthimer$^{3,5}$
\\
$^{1}$Department of Astronomy, University of Illinois Urbana-Champaign, Urbana, IL 61801, USA\\
$^{2}$Department of Astronomy, University of California, Berkeley, Berkeley, CA 94720, USA\\
$^{3}$Radio Astronomy Laboratory, University of California, Berkeley, Berkeley, CA 94720, USA\\
$^{4}$Theoretical Astrophysics Center, University of California, Berkeley, Berkeley, CA 94720, USA\\
$^{5}$Space Sciences Laboratory, University of California, Berkeley, Berkeley, CA 94720, USA
}

\date{Accepted XXX. Received YYY; in original form ZZZ}

\pubyear{\the\year{}}

\begin{document}
\label{firstpage}
\pagerange{\pageref{firstpage}--\pageref{lastpage}}
\maketitle

% Abstract of the paper
\begin{abstract}
The Long-Integration Magnetar Burst Observatory (LIMBO) is a real-time radio transient detection pipeline designed to search for dispersed fast radio bursts (FRBs) from Galactic magnetars. LIMBO employs a \leuschsize dish with a dual-polarization feed to continuously monitor a $125~\mathrm{MHz}$ effective band centred at $1460~\mathrm{MHz}$. A real-time processing pipeline performs a search for dispersed transients on the summed polarizations, with detections triggering dumps of buffered voltage data to disk. Based on calibrated sensitivity measurements, synthetic signal-injection and recovery tests, and successful detection of pulses from the Crab Pulsar, we determine that LIMBO is sensitive to radio transients with fluences $\gtrsim43~\mathrm{Jy \cdot ms}$. Between May and August 2023, LIMBO conducted 833~hours of follow-up observations of the Galactic magnetar \sgr, yielding 12 candidate FRB detections. If these events are true, we measure FRB-like event rates from \sgr of $R(\geq65~\mathrm{Jy \cdot ms}) = 112.3^{+81.3}_{-54.5}~\mathrm{yr}^{-1}$ and $R(\geq 130~\mathrm{Jy \cdot ms}) = 17.7^{+40.8}_{-15.1}~\mathrm{yr}^{-1}$. Combining these results with previously reported FRBs-like from \sgr, we infer a cumulative rate–fluence power-law slope of $\alpha=-0.60^{+0.24}_{-0.28}$ in the fluence range between 10 and $10^6\rm\, Jy \cdot ms$. These observations demonstrate the capability of continuous, real-time monitoring of Galactic magnetars and establish LIMBO as an effective instrument for detecting Galactic FRBs.
\end{abstract}

% Include between one and six keywords.

%%%%%%%%%%%%%%%%%%%%%%%%%%%%%%%%%%%%%%%%%%%%%%%%%%

%%%%%%%%%%%%%%%%% BODY OF PAPER %%%%%%%%%%%%%%%%%%

\section{Introduction} \label{sec:intro}
Fast radio bursts (FRBs) are bright, millisecond-duration radio transient events.
The first FRB was identified in 2007 through retrospective analysis of archival data from the Parkes radio telescope \citep{Lorimer2007}. 
Since then, hundreds to thousands of FRBs have been detected, with dispersion measures (DMs) often significantly exceeding the Galactic contribution, confirming extragalactic and cosmological origins \citep{Thornton2013, Petroff2016, Cordes2019, Gordon2023, CHIME/FRB_CAT2}. 
While the majority of FRBs appear to be one-off events, a subset has been observed to repeat \citep{Spitler2016, CHIMEFRB2023repeaters}.

These discoveries have been driven primarily by wide-field, high-cadence survey instruments, with major contributions from Parkes \citep{Thornton2013, Champion2016, Bhandari2018}, CHIME \citep{CHIME2018, CHIME/FRB_CAT1, CHIME/FRB_CAT2}, ASKAP \citep{Bannister2017, Shannon2018}, UTMOST \citep{Bailes2017, Caleb2017}, FAST \citep{2018IMMag..19..112L, 2021ApJ...909L...8N}, STARE-2 \citep{STARE2_2020}, ATA \citep{ATA2009, Sheikh2024}, and Apertif \citep{Apertif2017, Oostrum2020}, among others.

Although FRBs were first identified more than a decade ago, their physical origin and emission mechanism remain poorly understood, making them a compelling focus of astrophysical research.
Despite the large number of detections to date, nearly all FRBs have been observed at cosmological distances, making it difficult to definitively identify their progenitors, local environments, or emission processes. 

Consequently, a wide range of progenitor models have been proposed \citep{PetroffHesselsLorimer2019, Platts2019}, including compact object mergers \citep{Totani2013}, the collapse of supramassive neutron stars ("blitzars"; \cite{FalckeRezzolla2014}), neutron star-black hole interactions \citep{Mingarelli2015}, young energetic pulsars producing supergiant pulses \citep{CordesWasserman2016}, and highly magnetized neutron stars known as magnetars \citep{PopovPostnov2007, Lyubarsky2014, Murase2016, Beloborodov2017}.
Among these, magnetars emerged as a highly favoured candidate due to their extreme magnetic fields and high-energy activity. 
Strong support for this progenitor model came in 2020 with the detection of a bright radio burst from the Galactic magnetar \sgr that was temporally coincident with an X-ray flare \citep{Bochenek2020, CHIME_SGR2020, Mereghetti2020, Ridnaia2020, Li2021}.
This event demonstrated that magnetars are capable of producing FRB-like emission, providing the first direct evidence that at least some FRBs originate from them.
This discovery opens a promising pathway for constraining FRB emission mechanisms through detailed multi-wavelength observations of nearby magnetars \citep{Pelliciari2023, Geminardi2025, Wharton2025}. 
Subsequent follow-up observations of \sgr have resulted in successful repeat FRB-like detections, though generally at much lower burst energies. \citep{FAST_SGR2020, Kirsten2020, FASTR_SGR2022, CHIME_SGR2022, GBT_SGR2022, Yunnan_SGR2022, CHIME_SGR2022b}.

\begin{figure*}
    \centering
    \includegraphics[width=0.85\linewidth]{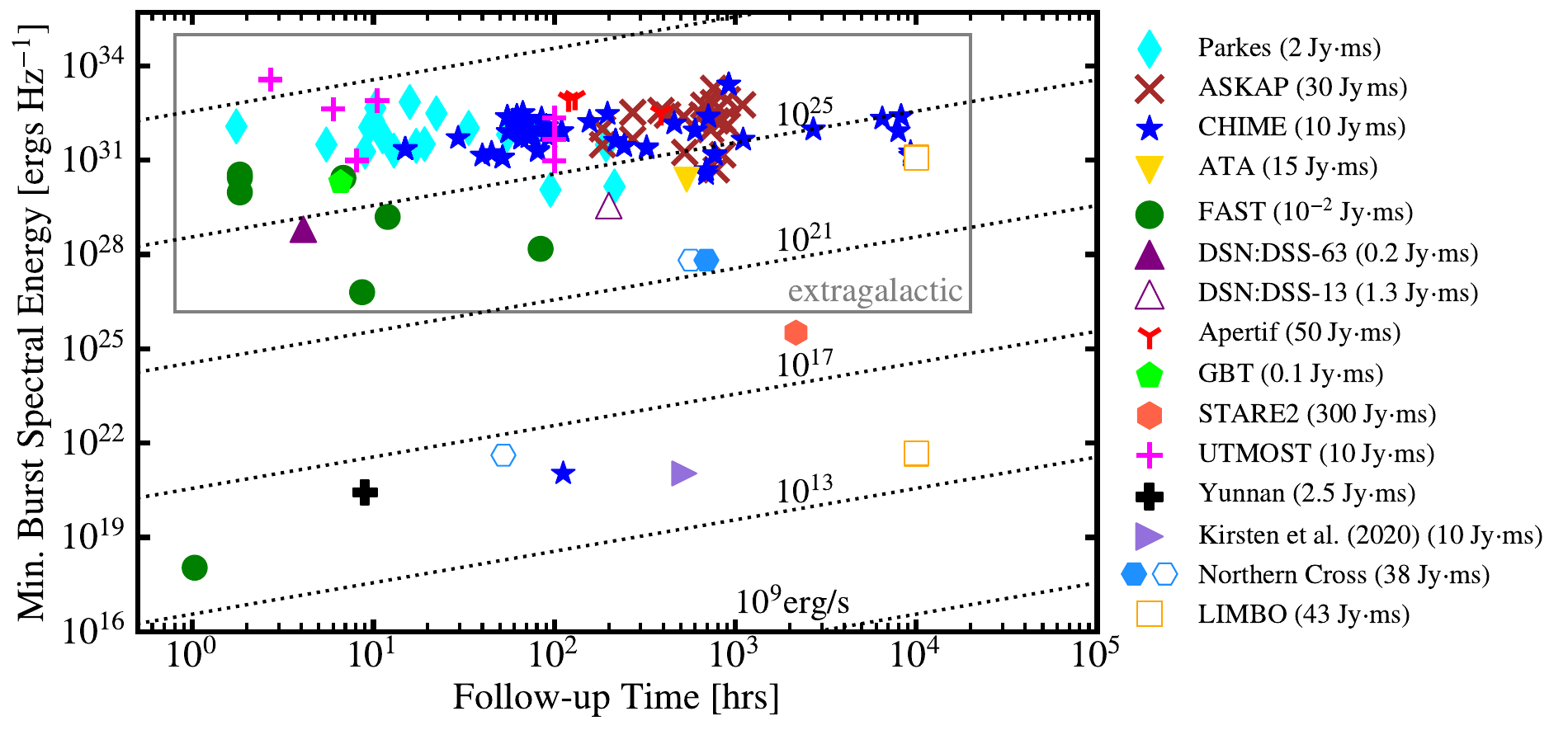}
    \caption{Minimum FRB burst spectral energy versus follow-up observation time for a range of FRB projects. For extragalactic sources, each symbol represents a different target observed by a given instrument; the symbols outside the "extragalactic" region correspond to searches for bursts from the Galactic magnetar \sgr using different instruments. Filled markers denote observing campaigns that detected one or more bursts, while open markers denote campaigns with no detected bursts and indicate the corresponding minimum detectable burst spectral energy based on the survey sensitivity. For comparison, we show LIMBO’s theoretical minimum detectable burst energy in both the Galactic and extragalactic (assumed redshift of $z=0.1$) regimes.}
    \label{fig:frb_followup}
\end{figure*}

Despite growing evidence linking at least a subset of FRBs to magnetars, the conditions under which magnetars produce these radio bursts remain poorly constrained. Proposed emission mechanisms are broadly divided between magnetospheric processes and relativistic shock models \citep{Kumar2017, 2020MNRAS.498.1397L, Metzger2019, Beloborodov2020, Margalit2020}. 
Extragalactic detections, however, are subject to strong distance-dependent sensitivity biases that favour the detection of high-energy bursts, sparse temporal coverage, and the inability to resolve progenitor sources via e.g., X-ray counterparts and identification of supernova remnants \citep{Cordes2019, PetroffHesselsLorimer2019, Hashimoto2022, Shin2023}.
This makes it difficult to characterize how the FRB emission is correlated with progenitor activity, although population modelling of the observed FRB luminosity and energy distributions, event rates, host-galaxy demographics, and redshift evolution provides important constraints on the underlying source population, likely progenitors, and emission mechanisms \citep{Eftekhari2017, Macquart2018, Zhang2021, Shin2023, Kirsten2024, Lei2025, Gordon2025, Jia2026}.

By contrast, long-term radio-frequency monitoring of Galactic magnetars for FRB activity offers a powerful approach to addressing these challenges. 
Targeting Galactic sources significantly reduces energy thresholds by orders of magnitude, making low-energy bursts observable. Furthermore, sustained, extended, and source-resolved follow-up observations of FRB-producing Galactic sources are critical for discriminating different FRB emission mechanisms.
In this way, dedicated single-dish facilities provide access to burst properties and temporal behaviours that are largely inaccessible to wide-field extragalactic surveys.

We present the Long-Integration Magnetar Burst Observatory (LIMBO), a real-time detection pipeline employed on the \leuschsize radio telescope at the University of California, Berkeley's Leuschner Observatory, optimized to perform long observational campaigns of Galactic magnetars and study the transient radio emission they produce. 
As illustrated in Figure \ref{fig:frb_followup}, a key strength of LIMBO is its capacity to accumulate $\sim10^4$ hours of follow-up observations, coupled with the sensitivity to detect faint Galactic FRBs, enabling exploration of a region of parameter space that has previously remained inaccessible. The figure also places LIMBO in the context of other FRB-detecting instruments reported in the literature \citep{Masui2015, Shannon2018, Farah2019, Zhu2020, Luo2020, Bochenek2020, CHIME_SGR2020, Connor2020, Niu2021, Kirsten2020, Majid2021, FASTR_SGR2022, Yunnan_SGR2022, Pelliciari2023, Sheikh2024, Wharton2025, Geminardi2025}. 

We organize the paper as follows. 
In Section \ref{sec:observering_system}, we describe the LIMBO instrument, including both its analogue and digital subsystems. 
Section \ref{sec:analysis} details calibration procedures, radio-frequency interference (RFI) excision techniques, and the real-time radio transient detection pipeline. 
In Section \ref{sec:commissioning}, we demonstrate signal-injection/recovery tests and detection of giant pulses from the Crab Pulsar. 
Observational targets and candidate FRB detections from \sgr using LIMBO are presented alongside inferred rate constraints in Section \ref{sec:tentative frb detections}.
Finally, we summarize results in Section \ref{sec:conclusions} and discuss the role of long-term, single-dish follow-up observations in advancing our understanding of FRBs.

\section{The Observing System} \label{sec:observering_system}

\begin{figure}
    \centering
    \includegraphics[width=0.95\linewidth]{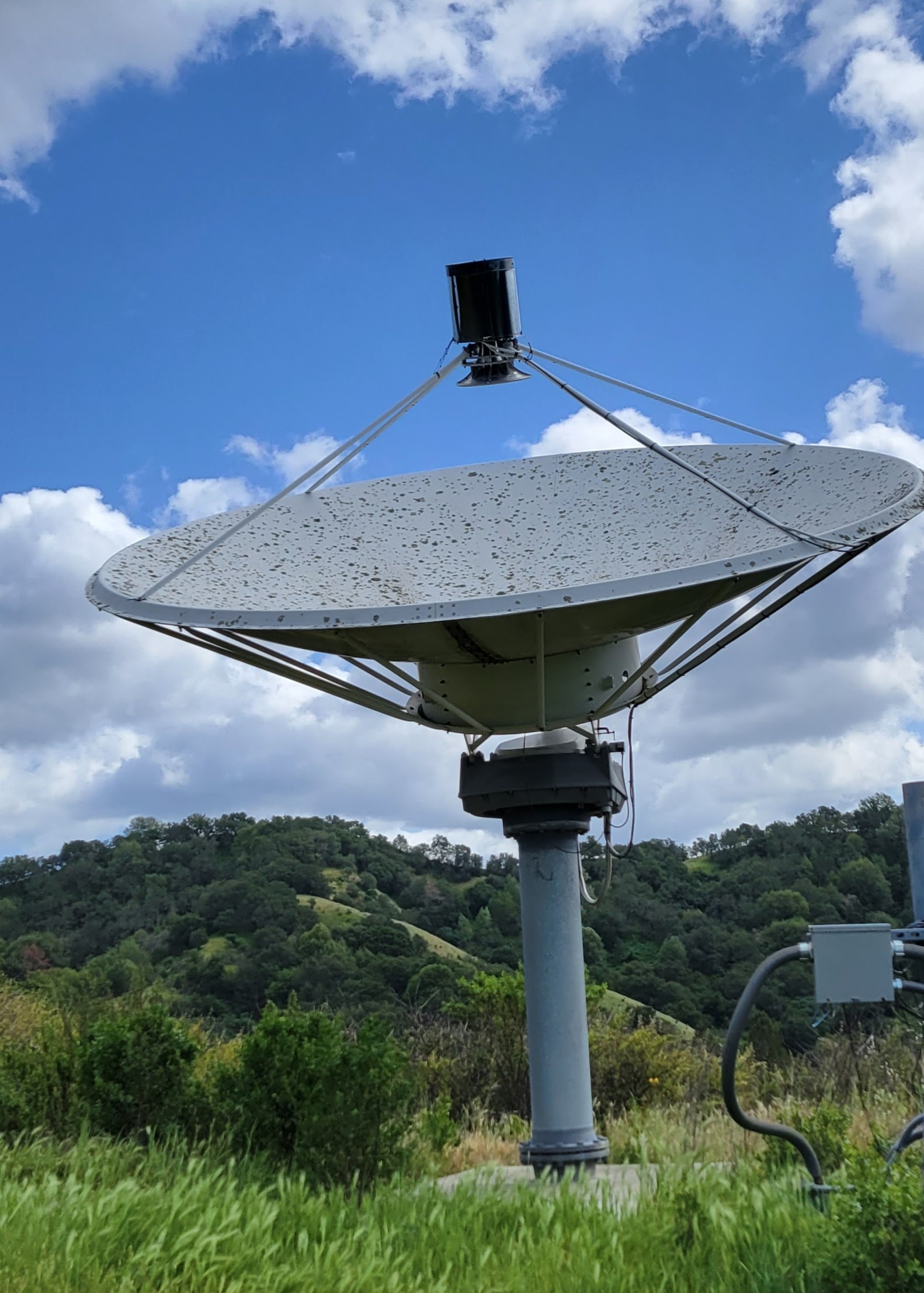}
    \caption{The \leuschsize Leuschner Radio Telescope located at the UC Berkeley Leuschner Student Observatory, equipped with the LIMBO FRB detection pipeline.}
    \label{fig:leuschner_dish}
\end{figure}

LIMBO operates with a \leuschsize radio dish (hereafter referred to as the Leuschner Radio Telescope) at the University of California, Berkeley’s Leuschner Student Observatory, located at $\left(-122^\circ~9' 12.31'', +37.8732^\circ \right)$. 
The dish, mounted on an alt-az drive, covers a declination range of $53^\circ$ to $123^\circ$, with a beam solid angle of $\sim 0.003~\rm sr$ and effective bandwidth of $125~\rm MHz$ (see Table \ref{tab:telescope_properties} for additional specifications and Figure \ref{fig:leuschner_dish}). 

\subsection{The Analogue System} \label{sec:analog_system}

\begin{figure*}
    \centering    
    \includegraphics[scale=0.17]{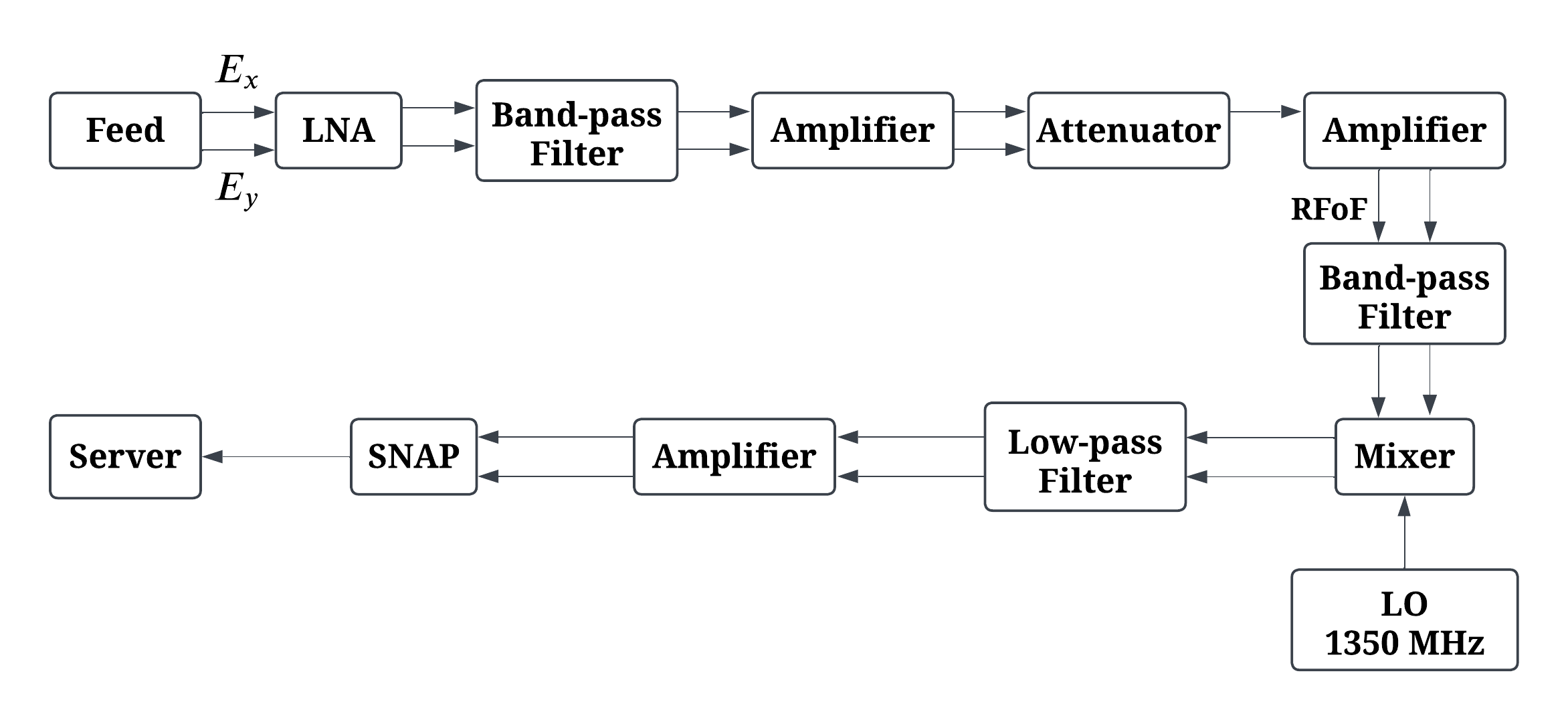}
    \caption{Schematic of the LIMBO analogue system. A dual-polarization feed horn captures signals which are amplified and filtered before being transmitted to an indoor processing rack via fibre-optic cabling. Inside, signals are heterodyne mixed, filtered, and further amplified before being digitally sampled on a SNAP board and sent to a local server over a 10-Gb Ethernet connection. The two linear polarizations follow parallel, matched paths on their way to the server.}
    \label{fig:analog}
\end{figure*}

The Leuschner Radio Telescope is equipped with a dual-polarization, wide-band, quadruple-ridged, flared horn feed \citep{2017ITAP...65.7322S}, uncooled low-noise amplifiers (LNAs), and radio-frequency-over-fibre (RFoF) modules that transmit signals over a 100-m link to an indoor processing rack for further amplification, down-conversion, filtering, and digital sampling. 
A block diagram of the LIMBO system is shown in Figure \ref{fig:analog}.

LIMBO processes two linear polarization streams through parallel analogue signal chains with matched characteristics. 
Each signal first passes through an LNA covering $0.8$–$2.5$~GHz, after which it is bandpass filtered to $1.2$–$2.5$~GHz. 
Additional amplification is provided by Mini-Circuits ZJL-7G+ and ZVA-183+ modules, with anti-reflection attenuation, before transmission via fibre-optic link using a Photonic Systems Inc. transmitter (PSI-1600-10UT) and receiver (PSI-1600-10UR).

Within the observatory’s control room, signals are filtered by a $1515 \pm 115$~MHz bandpass filter, then heterodyne mixed to baseband using an Agilent N9310A RF signal generator set to 1350~MHz as the local oscillator (LO).
This yields a total observable RF range of 1350-1600~MHz (250 MHz bandwidth).

\subsection{The Digital System} \label{sec:digital_system}

Analogue signals are digitized and packetized using a Smart Network ADC Processor (SNAP\footnote{\url{https://casper.astro.berkeley.edu/wiki/SNAP}}) board developed by the Collaboration for Astronomy Signal Processing and Electronics Research (CASPER) \citep{2017PASP..129d5001D, 2016JAI.....541001H}. 
Each of two 8-bit analogue-to-digital converters (ADCs) is clocked at 500~MHz, providing a Nyquist bandwidth of 250~MHz per polarization stream. 
The digitized signals are routed to the board’s Kintex-7 field-programmable gate array (FPGA), which performs real-time digital signal processing.

\begin{figure}
    \centering
    \includegraphics[scale=0.60]{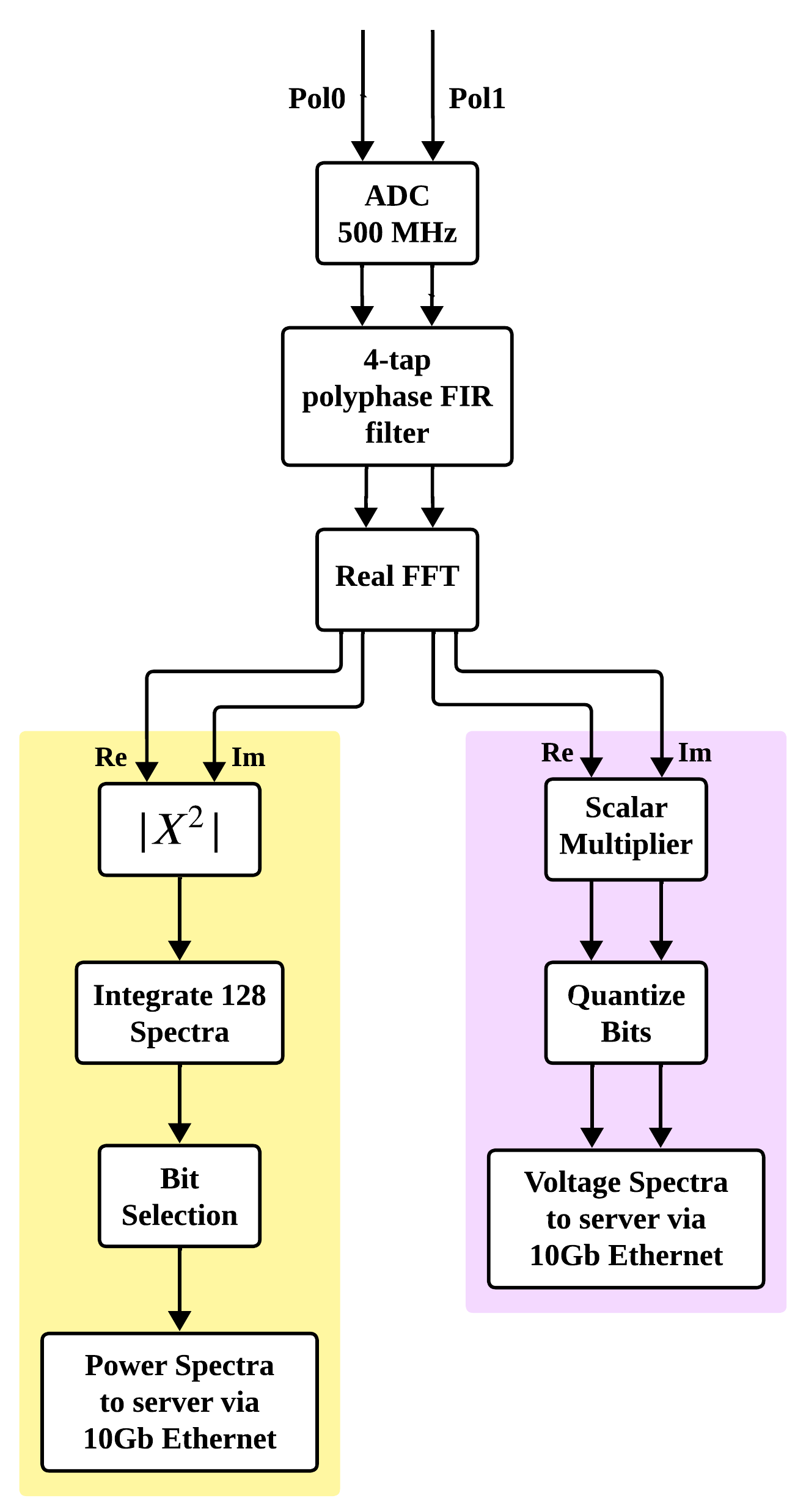}
    \caption{Schematic of the LIMBO digital signal processing system. Analogue signals are digitized by dual 8-bit, 500~MHz ADCs on a SNAP board and processed in real time by the onboard Kintex-7 FPGA, the design of which is based upon that of the DSA-10 \citep{2019MNRAS.489..919K}. A polyphase filter bank converts time-domain data into 2048 frequency channels, which are split into parallel power and voltage streams. Power spectra are integrated and down-sampled before transmission to a local server, while voltage spectra are quantized and streamed with high time resolution. Both streams use 10-Gb Ethernet with the HASHPIPE protocol.}
    \label{fig:digital}
\end{figure}

LIMBO employs a modified version of the DSA-10 SNAP design \citep{2019MNRAS.489..919K}, adapted to accommodate a different number of ADC inputs. 
FPGA firmware is designed using CASPER libraries. 
A polyphase filter bank (PFB), consisting of a 4-tap finite impulse response (FIR) filter followed by a 4096-point real-valued fast Fourier transform (FFT), channelizes the time-domain data into the frequency domain \citep{2009ISPL...16..477P}.
This produces 2048 frequency channels with a spectral resolution of 122~kHz.
The PFB output is then divided into two parallel spectral streams: a power stream and a voltage stream, each retaining 18-bit real and imaginary components.
A block diagram of the LIMBO digital system is shown in Figure \ref{fig:digital}.

Power spectra are integrated over 128 consecutive time windows, yielding a time resolution of 1~ms.
The data are then rescaled to signed 16-bit integers before being transmitted to the server over a 10-Gb Ethernet connection.
Similarly, voltage spectra from both polarizations are rescaled and quantized to signed 8-bit (4b real, 4b imaginary) complex integer format before being streamed to the server via a second 10-Gb Ethernet connection.

Both Ethernet streams employ the HASHPIPE data transfer protocol\footnote{\url{https://github.com/david-macmahon/hashpipe.git}}. 
The voltage spectra have a native time resolution of 8.192~$\mu$s and an aggregate data rate of $\sim$4~Gbps. 
These data are written to a 16-GB memory-mapped ring buffer on a CPU server, where they are retained for up to 16~s.
During this interval, real-time analysis and statistical thresholding are applied to the integrated power spectra to search for dispersed transient signals, as described below.

\section{LIMBO Analysis} \label{sec:analysis}

The LIMBO real-time analysis system operates on integrated power spectra in 4-s blocks to decide whether voltage spectra stored in the ring buffer should be overwritten or offloaded to long-term storage for follow-up analysis. 
The objective is to identify potential dispersed pulses with a negligible false-negative rate, while maintaining a false-positive rate low enough to minimize unnecessary data transfers that could delay processing and inflate storage and post-processing costs.

The real-time system incorporates gain and phase calibration, RFI excision, de-trending, statistical characterization, and an initial detection pipeline. 
Each of these components is described in detail below.

\subsection{Calibration} \label{sec:calibrations}

To obtain accurate measurements of flux density and polarization, we perform both a gain calibration and a polarization phase alignment. 
At LIMBO’s observing frequencies, the measured sky signal is dominated by diffuse Galactic synchrotron emission. 
As a result, observations of a well-characterized calibrator source provide a robust means of isolating the instrumental response and characterizing the frequency-dependent system temperature. 
Paired observations of Cygnus~A and a nearby cold-sky position ($20\rm h~10\rm m~30.44\rm s$, $+40^\circ~43'~34.31''$) are used to derive a frequency-dependent gain factor, enabling the conversion of measured power into absolute flux density \citep{1977A&A....61...99B}. 
To account for residual flux from Cygnus~A remaining within the beam, we model the beam as Gaussian and explicitly incorporate the resulting contamination into the derived gain factor.

To estimate the receiver temperature $T_\mathrm{rx}$, we account for the sky contribution using the Global Sky Model 2016 \citep{GSM2016}. 
We convolve Leuschner's beam with the all-sky continuum map to determine the expected sky temperature at 1420~MHz.
Centring the beam on the coordinates of \sgr, we measure an antenna temperature of 5.68~K for the local sky environment. Subtracting this contribution from the total system temperature $T_\mathrm{sys}$, we obtain a receiver temperature of 47.34~K at 1420~MHz.
This and other properties of the Leuschner Radio Telescope are summarized in Table \ref{tab:telescope_properties}.

\begin{table}
    \centering
    \caption{Leuschner radio telescope properties. The effective bandwidth describes the frequency channels used in the initial detection pipeline rather than the total observable frequency band (see Section \ref{sec:RFI}). The fluence sensitivity represents the minimum detectable fluence at the Z-score trigger threshold of 5.5, calculated using the measured system temperature and RMS noise temperature over the effective bandwidth and the calibrated gain (K/Jy; see Section \ref{sec:initial_detection}).}
    \begin{tabular}{c c}
        \hline \hline
        \textbf{Parameter} & \textbf{Value}\\
        \hline
        Beam Width & 3.43 degrees \\
        Effective Bandwidth & 125 MHz \\
        T$_{\mathrm{rx}}$, 1420 MHz & 47.34 K \\
        K/Jy & $5.73\times10^{-3}$ \\
        Sensitivity & 42.2 Jy$\cdot$ms \\
        \hline
    \end{tabular}
    \label{tab:telescope_properties}
\end{table}

For phase calibration, we use injected electromagnetic pulses to measure the relative phase difference between the two voltage streams (see Section \ref{sec:signal_injection} for further details). 
By correlating the injected signals, we determine the delay offset, which is corrected by applying a complex per-frequency phase adjustment to one of the streams. 
This procedure yields the final phase calibration.

\subsection{RFI Excision} \label{sec:RFI}

One of the primary challenges faced by LIMBO is the high level of RFI contamination. 
The Leuschner Radio Telescope’s location in the greater San Francisco Bay Area makes it susceptible to both narrow and broadband interference. 
Effective mitigation therefore requires careful modelling of the passband and robust excision strategies that suppress RFI "hot spots" while minimizing the risk of inadvertently flagging genuine FRB signals. 
Although LIMBO’s nominal passband extends from 1350--1600~MHz, we restrict our analysis to the cleaner 1398--1523~MHz range, excluding frequency regions persistently contaminated by RFI, for an effective bandwidth of $125~\rm MHz$. 
Bandpass modelling, RFI excision, initial detections, and FRB searches are therefore performed only within this restricted band.

\begin{figure}
    \centering
    \includegraphics[width=0.95\linewidth]{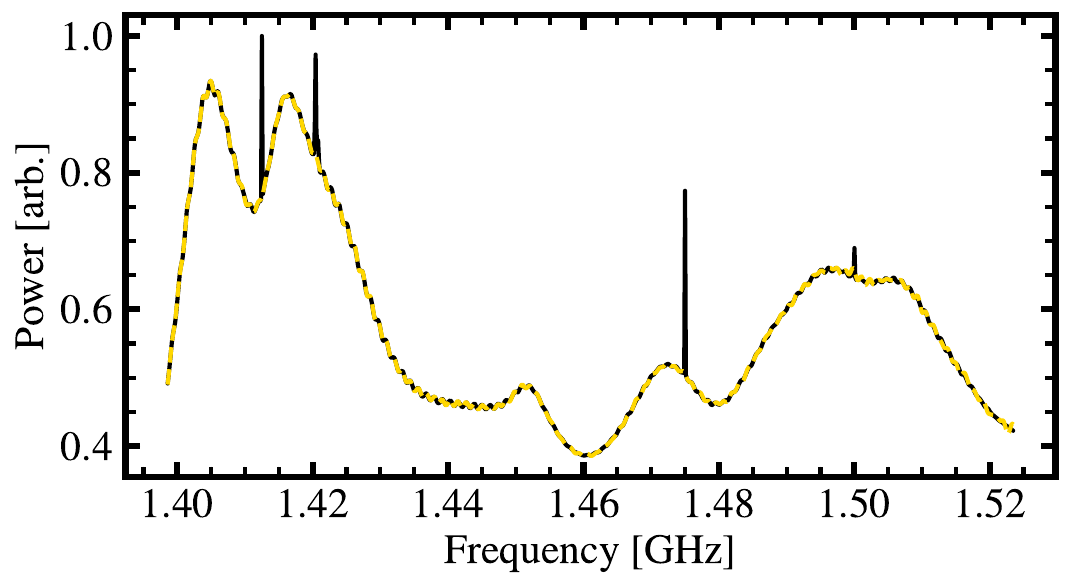}
    \caption{Modelling of Leuschner's bandpass using DPSS filters. The black curve is an uncalibrated Cygnus A spectrum with arbitrary units, time-averaged over $\sim28$~s. A DPSS filter excises clock-lines, RFI, and narrow spectral features, yielding a smoothed passband model (dashed yellow). }
    \label{fig:bandpass_modeling}
\end{figure}

To model the LIMBO passband, we use DPSS Approximate lazY filtEriNg of foregroUnds (DAYENU) \citep{2021MNRAS.500.5195E}. 
Applied to a time-averaged spectrum, this filter produces a smooth spectral model of the passband. 
As illustrated in Figure \ref{fig:bandpass_modeling}, the DPSS filter excludes clock lines, RFI spikes, and other narrow spectral features, yielding a clean model of the underlying passband response.

For each data file, smoothed estimates of $T_\mathrm{sys}$ are derived by applying DPSS filters in time and frequency to auto-correlation spectra. 
From these, thermal noise estimates for each channel are modelled according to the radiometer equation, 
$$ T_{\mathrm{rms}} = \frac{T_\mathrm{sys}}{\sqrt{\Delta \nu \tau}} $$
where $T_\mathrm{rms}$ is the residual (root-mean-squared) noise temperature, $T_\mathrm{sys}$ is the (frequency-dependent) system temperature, $\Delta \nu$ is the width of a frequency channel, and $\tau$ is the integration time. 

For each time–frequency bin, we compute a Z-score relative to this smooth noise model. 
We then sum the Z-scores across frequency to obtain an aggregate Z-score as a function of time. 
After subtracting the median, we compute the median absolute deviation (MAD) of the distribution and flag outlying time integrations using a $4\times$ MAD threshold.

\begin{figure*}
    \centering
    \includegraphics[scale=0.40]{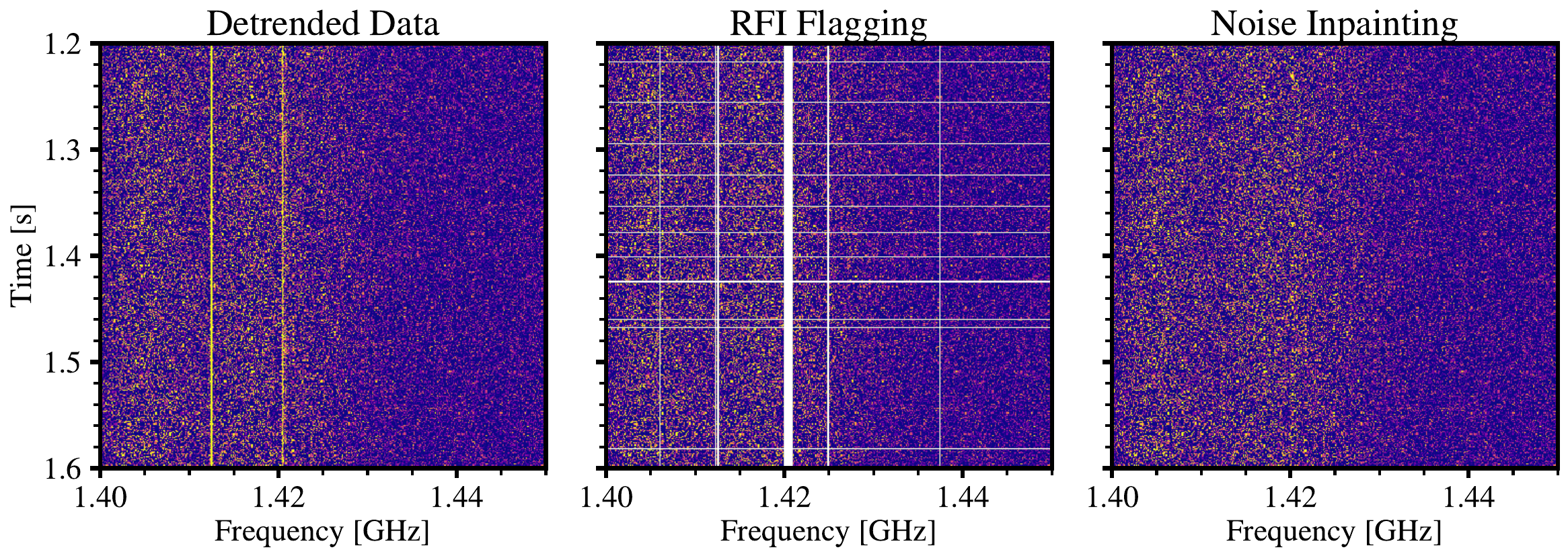}
    \caption{The central components of the LIMBO RFI excision pipeline, zoomed in time and frequency range for visual clarity. The left panel shows the data after it has been detrended according to our time and frequency-dependent models. The central panel shows, in white, all the time and frequency channels that were flagged for removal. The right panel contains inpainted Gaussian noise where time and frequency channels were removed.}
    \label{fig:rfi_flagging}
\end{figure*}

This procedure flags short bursts of excess power while preserving highly dispersed pulses, which distribute power over longer durations and therefore elevate both the median Z-score and MAD sufficiently to avoid rejection, even when per-bin Z-scores are large. 
Finally, after flagging time–frequency bins, we inpaint flagged data with Gaussian noise consistent with the frequency-dependent noise model to produce detrended, RFI-excised output data. We choose statistically consistent inpainting rather than zeroing flagged channels, as zeroing would artificially reduce the variance of the dedispersed time streams, biasing the event Z-scores to higher values, and inflating the number of false-positive detections.
The detrending, flagging, and inpainting process is illustrated in Figure \ref{fig:rfi_flagging}.

\subsection{Initial Detection: Searching for FRBs} \label{sec:initial_detection}

\begin{figure}
    \centering
    \includegraphics[width=0.85\linewidth]{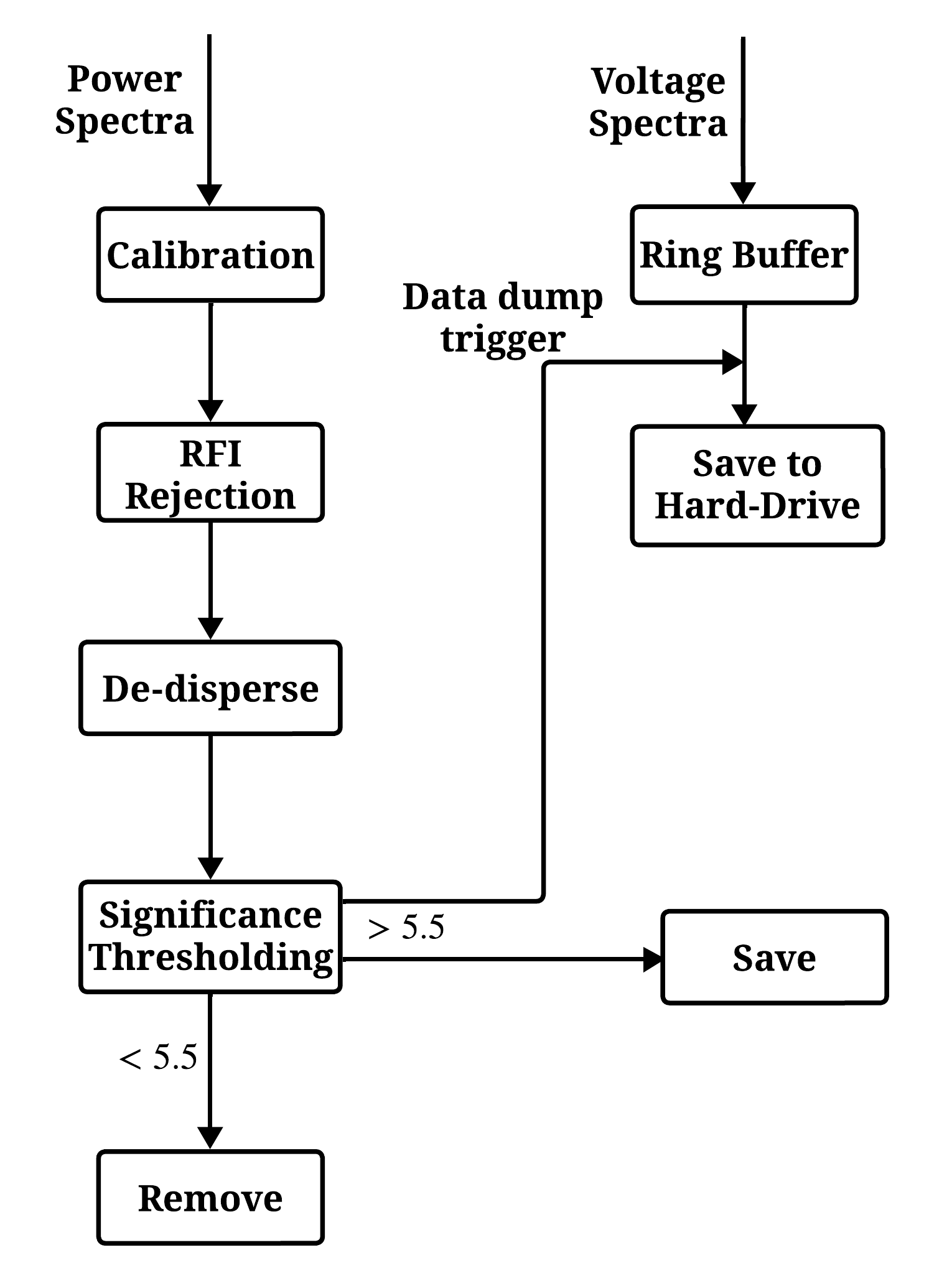}
    \caption{LIMBO's detection pipeline. The SNAP board outputs both voltage and power spectra. Initial detections are done using power spectra (left-hand side) using our real-time processing notebook template, which calibrates and dedisperses the data before searching for candidates. On the right side, voltage data is buffered until a detection is made, at which point both the power and voltage data are saved to hard-drive.}
    \label{fig:analysis}
\end{figure}

The initial detection pipeline is illustrated in Figure \ref{fig:analysis}.
While both power and voltage spectra are recorded simultaneously, initial detections are performed exclusively using power spectra; each power spectrum file is processed by a Jupyter notebook–based processing template that de-trends, calibrates, flags RFI, inpaints, and dedisperses the spectra.
The notebook generates diagnostic plots that allow the observer to quickly evaluate whether a file contains candidate pulses.
Whenever a new power spectrum file is written, a new notebook is automatically generated and executed to process that individual file.
The total processing time per file is shorter than the file buffering time, enabling real-time detection.

For sources with known DMs, LIMBO performs dedispersion directly at the reported values. 
For sources with unknown DMs, LIMBO can also conduct blind DM searches using a custom implementation of the Fast Dispersion Measure Transform (FDMT) algorithm \citep{ZackayOfek2017}. 
Implemented in Cython, FDMT$_{\mathrm{LIMBO}}$ is a fully coherent dedispersion algorithm with computational complexity $N_t N_f \log(N_t)\log(N_f)$, where $N_t$ and $N_f$ denote the number of time and frequency bins, respectively.
Planned extragalactic searches for FRBs in nearby galaxies and galaxies hosting known FRBs makes FDMT$_{\mathrm{LIMBO}}$ a valuable addition to the system.

\begin{figure}
    \centering
    \includegraphics[width=0.95\linewidth]{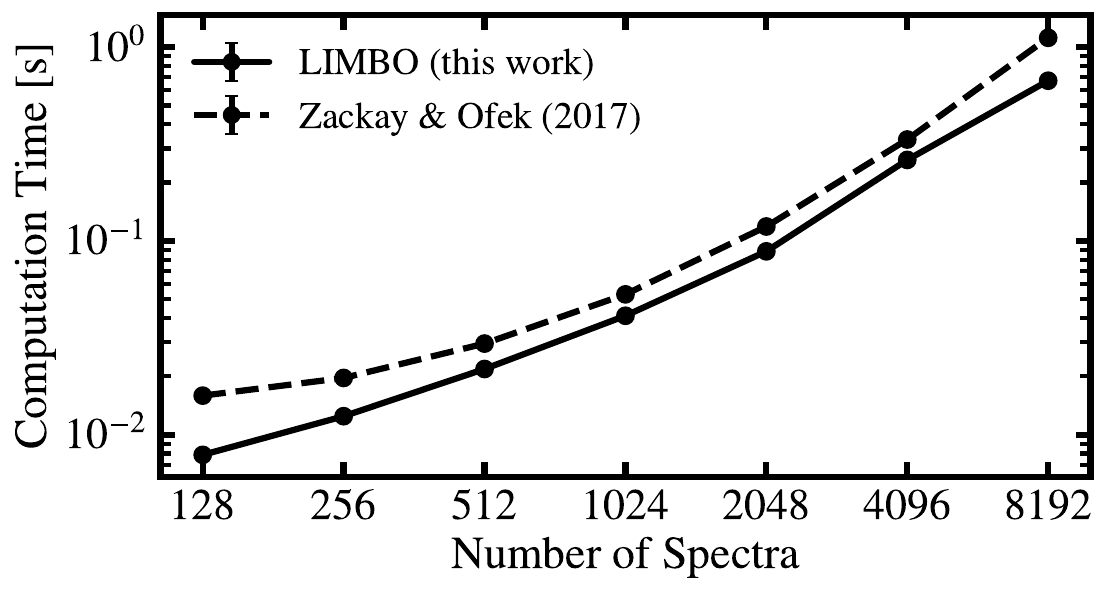}
    \caption{A comparison of the original \protect\cite{ZackayOfek2017} FDMT algorithm's run-times versus that of FDMT$_{\mathrm{LIMBO}}$. Each version was run on a 4.05~GHz Apple M3 processor. Varied as a function of the number of spectra, each dataset contained a constant 2048 frequency channels. The points correspond to the average run-time measured when sampled 100 times. The error bars indicate the standard deviation on that sample. These run-times show that the LIMBO implementation of the FDMT algorithm can dedisperse data up to $2\times$ faster than the original 2017 implementation, regardless of the number of spectra.}
    \label{fig:runtimes}
\end{figure}

\begin{figure*}
    \centering
    \includegraphics[width=0.90\linewidth]{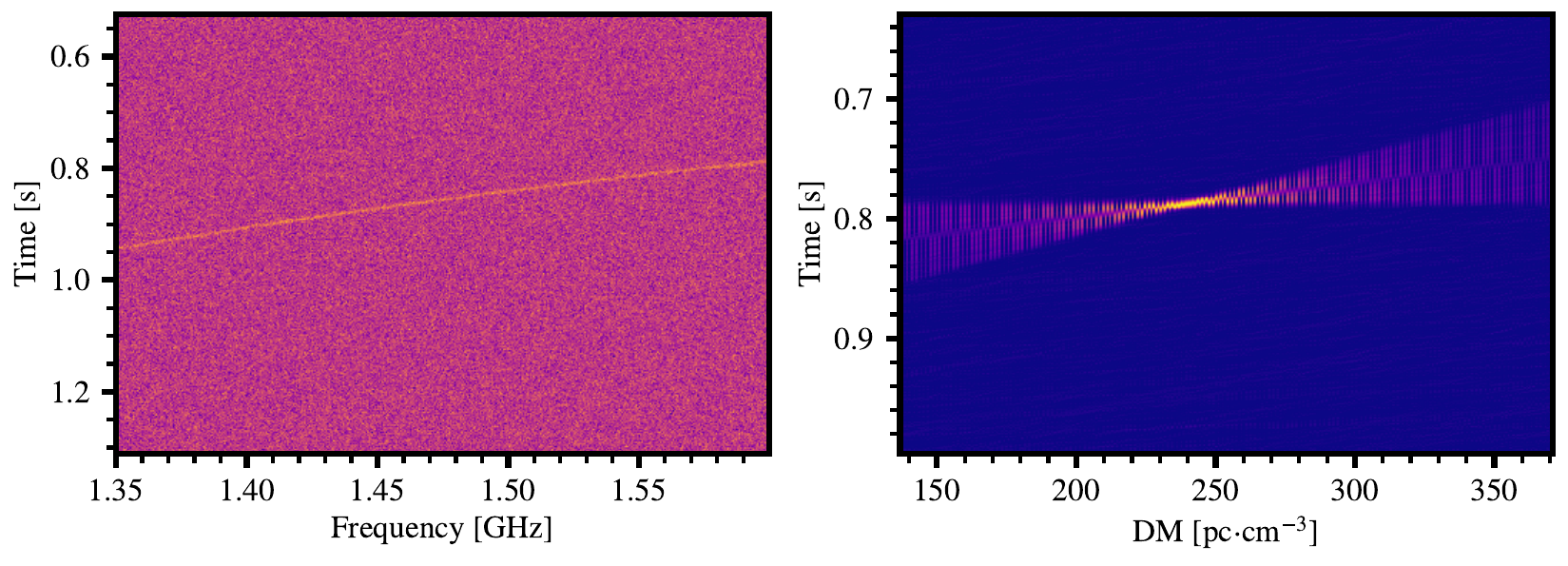}
    \caption{\emph{Left:} A computer simulated, dispersed pulse of width 2.12 ms and dispersion measure of 240 $\mathrm{pc/cm^3}$ sweeping across the LIMBO observing band. A power spectra as a function of time and frequency serves as the input to the FDMT algorithm. \emph{Right:} The output of the FDMT$_{\mathrm{LIMBO}}$ algorithm. The dispersion measure transform matrix is a function of dispersion measure and time, where the maximum reads the measured DM (in this case, 240.82 $\mathrm{pc/cm^3}$) and the time when the pulse enters the band.}
    \label{fig:fdmt_limbo}
\end{figure*}

We find that FDMT$_{\mathrm{LIMBO}}$ runs approximately twice as fast as the original implementation (Figure \ref{fig:runtimes}) on a 4.05~GHz Apple M3 processor, enabling real-time coherent dedispersion of candidate pulses without GPU acceleration. 
For a data matrix with 2048 frequency channels and 4096 spectra---typical of LIMBO power spectrum files---the algorithm completes a full dedispersion in $\sim$0.26~s and outputs the measured DM of any detected pulses. 
An example input and output of FDMT$_{\mathrm{LIMBO}}$ is shown in Figure \ref{fig:fdmt_limbo}.

Dedispersed events exceeding a Z-score significance threshold of 5.5 trigger a file-save response, prompting data dumps from the voltage ring buffer.
The Z-score is used as the triggering statistic because it characterizes the significance of a candidate relative to the empirical distribution of noise events produced by the full search pipeline, including non-Gaussian noise contributions, residual unflagged, contaminating RFI, and the large number of trials inherent to a blind transient search.
The threshold of 5.5 was selected as the lowest value that avoids overwhelming the system with spurious candidate pulses.

Using our measured $T_{\rm sys}$ and $T_{\rm rms}$ over the effective 125 MHz bandwidth, together with the calibrated gain (K/Jy; Table \ref{tab:telescope_properties}), we determine a minimum detectable fluence of $42.2~\mathrm{Jy}\cdot\mathrm{ms}$ at the Z-score trigger threshold.
Each saved voltage dataset is subsequently reprocessed through a similar pipeline to confirm the pulse detection. 
The availability of voltage data enables more detailed analyses of confirmed FRB candidates.

\section{Commissioning} \label{sec:commissioning}

\subsection{Signal Injection} \label{sec:signal_injection}

During the development stages, we validated the LIMBO system using two signal-injection methods.
First, we injected synthetic FRB signal into simulated power spectrum files with noise properties matched to LIMBO observations. 
These tests verify the notebook-based initial detection pipeline and ensure that FDMT$_\mathrm{LIMBO}$ correctly recovers dispersed pulses over a range of dispersion measures and signal strengths.
Second, we performed end-to-end hardware tests by manually injecting dispersed radio signal into the Leuschner feedhorn with a transmitter antenna located $\sim100~\rm m$ away from the dish. 
Using a Raspberry Pi 4 pared with a Programmed Test Sources (PTS) frequency synthesizer, we generate continuous-wave signals, linear frequency sweeps, and quadratic frequency sweeps design to mimic a dispersed FRB pulse. 
These controlled hardware injections test the real-time detection pipeline, including triggered voltage dumps, and confirm that the system responds quickly enough for real-time observations and detections.
These injections were not used to characterize the telescope beam response or for polarization calibration.

\subsection{Giant Crab Pulses} \label{sec:crab_pulses}

In addition to controlled synthetic tests, we used the Crab Pulsar (PSR~B0531+21) as a natural calibration source to validate LIMBO’s performance during observing runs. 
Giant Crab Pulses (GCPs) mimic FRBs closely enough to test all stages of the system, including the initial detection pipeline, triggered voltage dumps, data recording, and calibration routines under realistic sky conditions.

\begin{figure}
    \centering
    \includegraphics[width=0.85\linewidth]{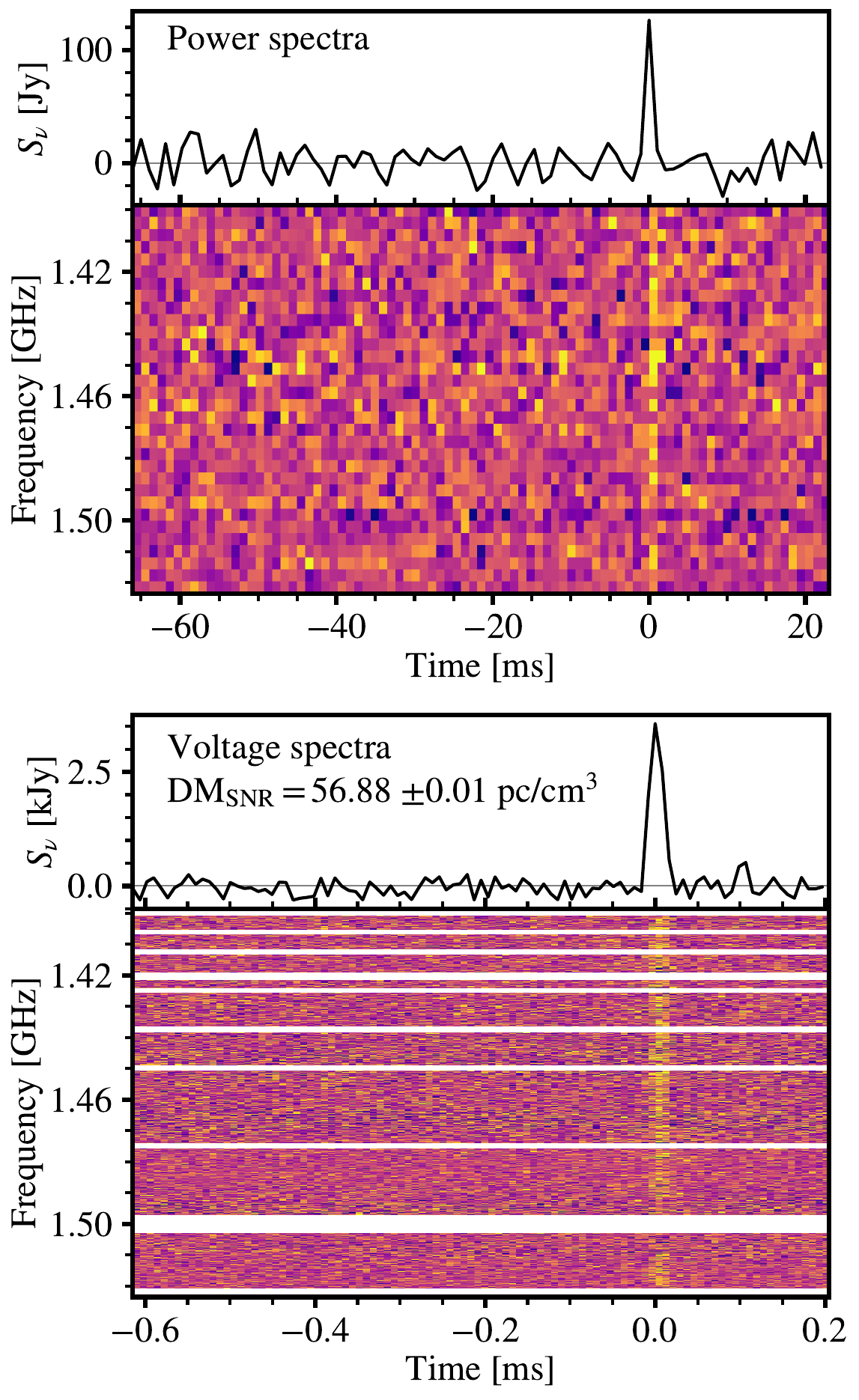}
    \caption{Detection of a giant radio pulse from the Crab Pulsar using LIMBO. \emph{Top:} The power spectra of GCP230707 as it is seen by LIMBO's initial detection pipeline. The pulse was dedispersed to the known DM of the Crab pulsar (56.7 pc$\cdot$cm$^{-3}$) for detection purposes. \emph{Bottom:} Corresponding voltage data of GCP230707. The pulse was dedispersed such that the SNR was maximized, giving a measured DM of 56.88~pc$\cdot$cm$^{-3}$. Flagged frequency channels are masked out in white. In each figure, the top panel shows the frequency-averaged flux density of the pulse as a function of time, while the bottom panel displays the time and frequency-domain profile. }
    \label{fig:GCP230707}
\end{figure}

An example detection is shown in Figure \ref{fig:GCP230707}, where LIMBO recorded a GCP with a measured dispersion measure of $56.88~\mathrm{pc\cdot cm}^{-3}$, consistent with the established value for the Crab Pulsar \citep{Lyne1993, McKee2018, Lewandowska2022}. 
Using the higher time resolution provided by the voltage data, we were able to resolve the pulse profile in detail. The event yielded a fluence of $66.8 \pm 6.3~\mathrm{kJy\cdot ms}$, a signal-to-noise ratio (SNR) of 21.46, and a pulse width of $32.77~\mathrm{\mu s}$, all consistent with published measurements of other GCPs \citep{Karuppusamy2010, Bera2019, Majid2011}.

These detections demonstrate that LIMBO can reliably capture and characterize dispersed astrophysical signals, validating the system performance in FRB searches.
While the recorded data could also be used for periodicity searches of pulsars, such analyses are beyond the scope of this work, which focuses exclusively on validating LIMBO for real-time FRB detections.

\section{Tentative FRB Detections} \label{sec:tentative frb detections}

\subsection{Targets} \label{sec:targets}

Using the McGill magnetar catalogue\footnote{\url{http://www.physics.mcgill.ca/~pulsar/magnetar/main.html}} \citep{McGill2014}, we determine that 22 of the 30 known Galactic magnetars are at a declination observable by the Leuschner Radio Observatory. 
The characteristics of these observable magnetars and the amount of time each of these sources are visible to the Leuschner Radio Observatory is given in Table \ref{tab:magentars}. 
This roughly represents the maximum amount of time each source could be targeted on any given day. 
This table, combined with reports of activity from any of these magnetars, informs our observing strategy.

\begin{table*}
\centering
\caption{22 of the 30 known Galactic magnetars are observable from the Leuschner Radio Observatory. Listed are their daily visibilities, dispersion measures (DMs), and characteristic ages.}
\label{tab:magentars}
    \begin{threeparttable}
        \begin{tabular*}{\textwidth}{@{\extracolsep{\fill}} l c c c c}
            \hline \hline
            \textbf{Source} &
            \textbf{LIMBO visibility [hrs/day]} &
            \textbf{DM [pc cm$^{-3}$]\tnote{a}} & 
            \textbf{$\tau_c$ [kyr]\tnote{b}} &
            \textbf{Ref.} \\ 
            \hline
            4U 0142+61 & $18.5$ & $27$ & $68$ & 1, 2 \\
            SGR 0418+5729 & $17.0$ & $50\,(5)$ & $3.6\times 10^4$ & 1, 2 \\
            SGR 0501+4516 & $14.4$ & \ldots & $15$ & 2 \\
            SGR J1745-2900 & $4.8$ &  $1778\,(3)$ & $4.3-9.7$ & 1, 2 \\
            SGR 1806-20 & $6.6$ & $\sim 750$ & $0.16-1.45$ & 2, 3 \\
            XTE J1810-197 & $6.8$ &  $178\,(5)$ & $8.4-20$ & 1, 2 \\
            Swift J1818.0-1607 & $7.4$ & $706\,(4)$ & $0.47$ & 1, 4 \\
            Swift J1822.3-1606 & $7.4$ & \ldots & $1600-6400$ & 2 \\
            SGR 1833-0832 & $8.4$ & \ldots & $34$ & 2 \\
            Swift J1834.9-0846 & $8.4$ & \ldots & $4.9$ & 2 \\
            1E 1841-045 & $8.9$ & $\sim 746$ & $4.5$ & 2, 3, 5 \\
            3XMM J185246.6+003317 & $9.5$ & \ldots &  $>1300$ & 2 \\
            SGR 1900+14 & $10.4$ & $281.4\,(0.9)$ & $0.46-1.3$ & 2, 6 \\
            SGR 1935+2154 & $11.6$ & $332.7206\,(9)$ & $3.6$ & 1, 2 \\
            1E 2259+586 & $17.4$ & $79$ & $230$ & 1, 2 \\
            SGR 0755-2933 & $4.6$ & \ldots &  \ldots & \ldots \\
            SGR 1801-23 & $6.2$ & \ldots &  \ldots & \ldots \\
            SGR 1808-20 & $6.6$ & \ldots &  \ldots & \ldots \\
            AX J1818.8-1559 & $7.4$ & \ldots & \ldots & \ldots \\
            AX J1845.0-0258 & $9.1$ & \ldots & \ldots & \ldots \\
            SGR 2013+34 & $13.0$ & \ldots &  \ldots & \ldots \\
            PSR J1846-0258 & $9.1$ & \ldots & $0.73$ & 2 \\ 
            \hline
        \end{tabular*}
    \begin{tablenotes}
        \footnotesize
        \item[a] {DM values are only populated when there is a published radio pulsar detection and a catalogued DM.}
        \item[b] {Characteristic (dipole spin-down) age.}
        \item[] {References: 
            (1)~\citet{McGill2014}; 
            (2)~\citet{Beniamini2019}; 
            (3)~\citet{Lazarus2012}; 
            (4)~\citet{Hu2020}; 
            (5)~\citet{Younes2025}; 
            (6)~\citet{Shitov1999}.}
    \end{tablenotes}
    \end{threeparttable}
\end{table*}
    
\subsection{Observations of \sgr} \label{sec:sgr}

Thus far, LIMBO observations have focused on the Galactic magnetar \sgr, which has produced a number of FRB-like bursts and has a well-characterized dispersion measure ($332.7~\rm pc\cdot 
\rm cm^{-3}$; \cite{Bochenek2020, FAST_SGR2020, Kirsten2020, FASTR_SGR2022, CHIME_SGR2022, Yunnan_SGR2022, CHIME_SGR2022b}). 
Between May and August 2023, we conducted a total of 833 hours of follow-up observations of this source.

Candidate FRBs in this dataset are identified through statistical thresholding combined with visual inspection of spectra. 
As outlined in Section \ref{sec:initial_detection}, we evaluate each file’s significance against a detection threshold to assess the likelihood of it containing an FRB. 
In the absence of systematics, the distribution of measured Z-scores would follow a standard normal distribution. 
In practice, however, imperfect RFI excision and residual systematics distort the distribution, shifting the mean toward positive values and inflating the reported significance. 

\begin{figure}
    \centering
    \includegraphics[width=\linewidth]{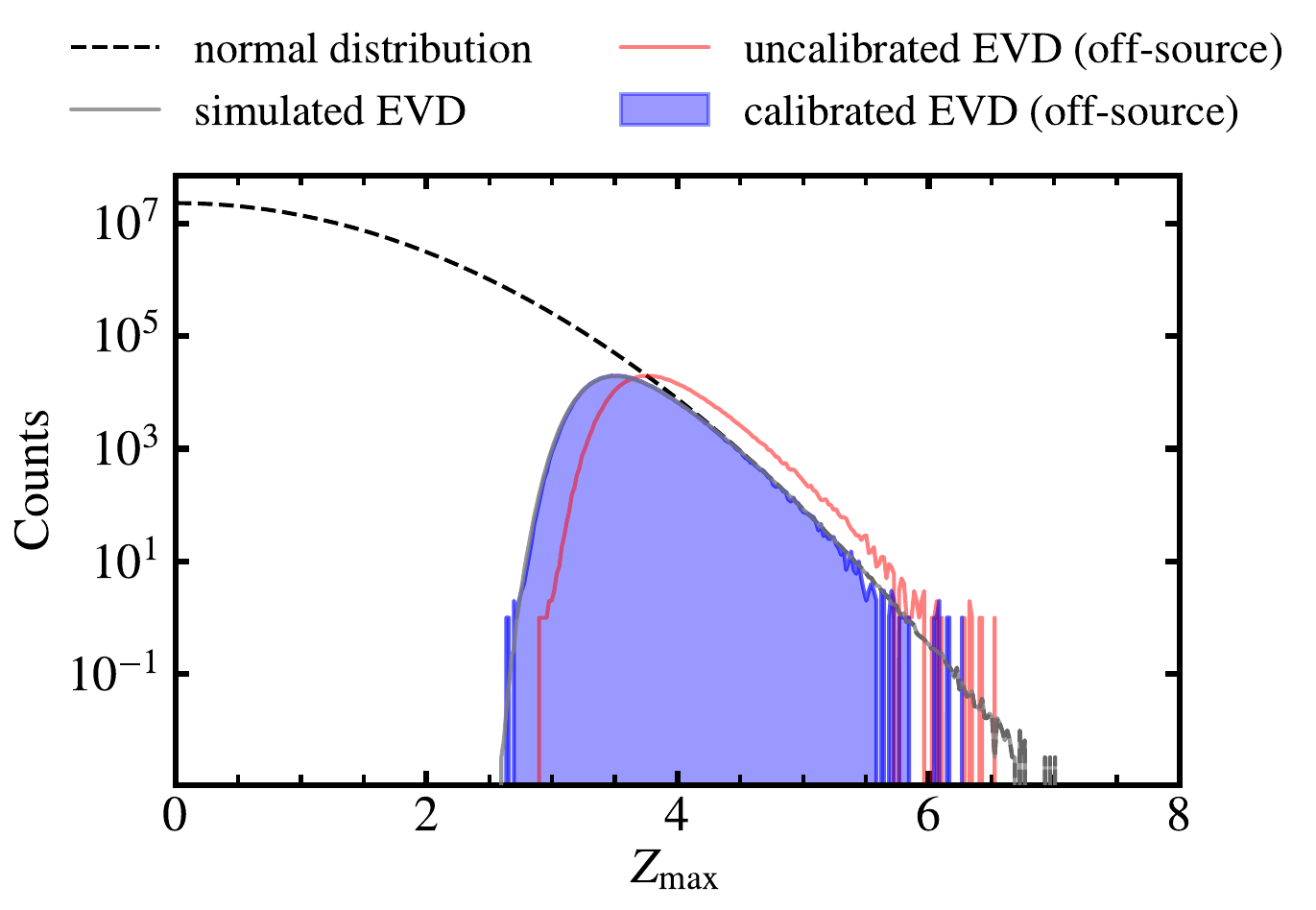}
    \caption{Histogrammed per-file maximum significances of an off-\sgr observing run plotted against the average of 100 simulated trials of pure Guassian noise. The red curve represents the uncalibrated EVD obtained from the observing run. Subtracting the difference between the means of the simulated EVD and uncalibrated EVD gives us a calibrated EVD for \sgr (blue). The calibrated EVD is compared to that expected by Gaussian noise (grey). Overplotted in black is a normal distribution, illustrating that the right tail of an EVD can be well approximated by a Gaussian distribution.}
    \label{fig:EVDs}
\end{figure}

Figure \ref{fig:EVDs} shows a histogram of the \emph{maximum} measured significance per LIMBO file, illustrating this bias.
We choose the most significant event within each file to evaluate as an FRB candidate. 
All detections are therefore evaluated against an extreme value distribution (EVD) expected from pure Gaussian noise.
Any systematic shift of the observed EVD relative to the simulated distribution reflects the combined effects of residual RFI and other unmitigated systematics.

To quantify this effect, we conducted a $\sim6~\mathrm{hr}$ calibration run slightly off-target from \sgr. 
This dataset characterizes the instrument’s triggering environment in the absence of genuine FRBs.
From this analysis, we find that unflagged RFI increases the measured Z-scores by an additive offset of approximately +0.25.
We therefore subtract this offset from all candidate events to obtain corrected significance values.

We then examine the shape of the calibrated EVD in greater detail. 
At high significance levels, the observed tail of the distribution is systematically suppressed compared to expectations, with events at $Z_\mathrm{max}>4.5$ occurring at rates $\sim10\%$ below the simulated Gaussian noise prediction. 
This deficit may be the result of over-flagging within the LIMBO RFI excision pipeline. 
However, comparisons of flagged and unflagged distributions show that only $\sim1\%$ of the suppression can be accounted for by differences in flagged channels, suggesting that a more detailed noise analysis and stronger characterization of the LIMBO pipeline is required. 
We acknowledge that this biases us towards under-reporting the significance of high-Z events, but for this work we do not attempt to apply additional corrections.

After applying an offset correction derived from the off-source calibration run, the measured EVD aligns much more closely with the simulations. 
This corrected Z-score is used for interpretation within the statistical framework assumed by our pipeline.

\begin{figure*}
    \centering
    \includegraphics[width=0.95\linewidth]{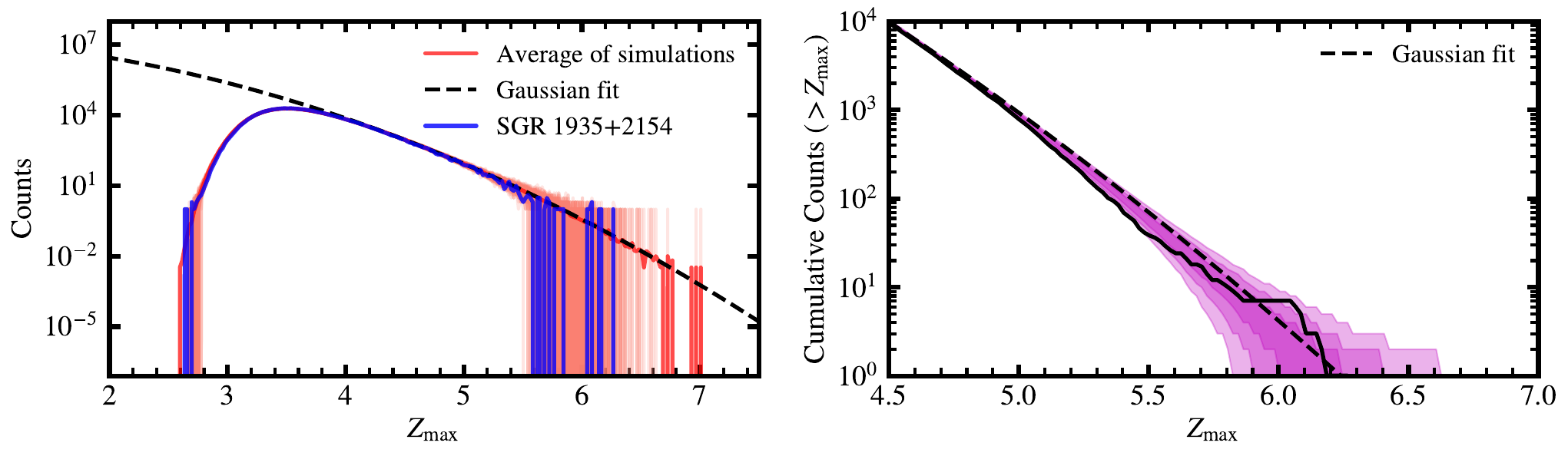}
    \caption{Histograms of the maximum Z-score observed per file within the May--August 2023 LIMBO observing campaign. \emph{Left:} EVDs of the \sgr observing run (blue) and 100 simulations (pink). The average EVD of these simulations is plotted in red. A Gaussian was fit to the tail of the averaged EVD (black dashed). \emph{Right:} A cumulative distribution of the \sgr observing run (black, solid) plotted against the same Gaussian fit to the average of the simulations in the left panel (black, dashed). The purple (from darkest to lightest) represent the 68\%, 95\%, and 99.7\% confidence intervals drawn from a a Poisson distribution sampled 300 times. The suppression of the distribution's tail from the Gaussian fit is the result of unmitigated systematics.}
    \label{fig:EVD stats}
\end{figure*}

By the central limit theorem, the right tail of an EVD is well-approximated by a Gaussian distribution, particularly at maximum Z-scores above the trigger threshold.
In Figure \ref{fig:EVD stats} (left), we directly compare the calibrated EVD of observations to an ensemble of simulations, while Figure \ref{fig:EVD stats} (right) presents the cumulative distribution of maximum Z-scores. 
A Gaussian is fit to the average of the simulated distributions. 
The cumulative plot reveals that the observed tail is suppressed relative to the Gaussian fit, with shaded regions marking the 68\%, 95\%, and 99.7\% confidence intervals derived from repeated Poisson resampling. 
Despite this suppression, we find a clear excess of events over the simulated expectation in bins with $Z_\mathrm{max} \geq 5.8$.

We select all events with significance $\geq 5.6$ for further analysis, yielding 24 candidate bursts. We expect false positives from Gaussian noise alone to be rare ($P(Z_\mathrm{max}\geq5.6)= 1\times10^{-8}$), making $Z_\mathrm{max} \geq 5.6$ a natural threshold for further examination of files.
Each candidate's power spectrum is visually inspected and compared to expectations for FRB-like signatures. 
Of these 24 events, we classify 12 as plausible FRBs, 7 as spurious events attributable to RFI, and 3 as a combination of RFI and FRB-like behaviour (see Appendices \ref{appx:candidates} and \ref{appx:rfis}). 
Candidate FRBs are presented in Figure \ref{fig:hits}, with pulse properties listed in Table \ref{tab:frb_props}. 
Events highlighted in bold represent the most secure identifications, supported by both high statistical significance and clean spectral-temporal profiles consistent with FRBs. 
We characterize the strength of each burst using it's SNR, which we use to determine the DM for each candidate by maximizing the burst SNR over a range of trial DMs centred on that expected for \sgr.
The 1$\sigma$ uncertainty on the measured DM is estimated by fitting a Gaussian to the resulting SNR--DM curve providing an estimate of the uncertainty arising from limited signal strength, intrinsic burst structure, and residual RFI contamination.
The measured DMs of these candidates are consistent with previously published values for \sgr \citep{Bochenek2020, FAST_SGR2020, Kirsten2020, FASTR_SGR2022, CHIME_SGR2022, Yunnan_SGR2022, CHIME_SGR2022b}.

\begin{table*}
    \centering
    \caption{Candidate FRBs detected by the LIMBO system between May and August 2023, along with their measured pulse properties. Here, the Z-score and SNR serve distinct purposes: the Z-score quantifies the statistical significance of a candidate within the empirical noise distribution of the search pipeline, while the SNR characterizes the strength of the burst signal and is used to determine the DM. Dispersion measures and their $1\sigma$ uncertainty estimates are obtained by maximizing the pulse SNR over a range of trial DMs and fitting a Gaussian to the resulting SNR--DM curve. Fluence uncertainties are calculated over a time window centred on the burst that is shorter than the full file duration, ensuring that possible RFI elsewhere in the file does not bias the noise statistics for the burst. Candidate names shown in bold indicate bursts judged to be most reliable after visual inspection of their power spectra.
    }
    \begin{tabular*}{\textwidth}{@{\extracolsep{\fill}} l c c c c}
        \hline \hline
        \multicolumn{1}{c}{\textbf{Name}} & \multicolumn{1}{c}{\textbf{DM [pc/cm$^3$]}} & \multicolumn{1}{c}{\textbf{Fluence [Jy$\cdot$ms]}} & \multicolumn{1}{c}{\textbf{SNR}} & \multicolumn{1}{c}{\textbf{Z-score}}
        \\ \hline
        \textbf{LIMBO230524A} & 330.5~$\pm$~1.3 & 79~$\pm$~15 & 6.12 & 5.85 \\
        \textbf{LIMBO230524B} & 329.4~$\pm$~1.8 & 80~$\pm$~13 & 6.19 & 5.91 \\
        \textbf{LIMBO230526} & 333.5~$\pm$~1.7 & 105~$\pm$~17 & 5.99 & 5.71 \\
        \textbf{LIMBO230527A} & 331.3~$\pm$~1.2 & 79~$\pm$~13 & 5.89 & 5.60 \\
        \textbf{LIMBO230527B} & 330.8~$\pm$~1.5 & 75~$\pm$~13 & 5.94 & 5.66 \\
        \textbf{LIMBO230608} & 332.1~$\pm$~1.7 & 78~$\pm$~13 & 6.04 & 5.76 \\
        \textbf{LIMBO230613} & 334.1~$\pm$~1.6 & 80~$\pm$~13 & 6.45 & 6.17 \\
        LIMBO230617 & 332.9~$\pm$~1.8 & 74~$\pm$~14 & 6.09 & 5.80 \\
        \textbf{LIMBO230708} & 334.1~$\pm$~1.5 & 82~$\pm$~13 & 6.51 & 6.21 \\
        \textbf{LIMBO230720} & 330.4~$\pm$~1.8 & 79~$\pm$~13 & 6.08 & 5.78 \\
        LIMBO230721 & 334.4~$\pm$~1.8 & 68~$\pm$~12 & 6.32 & 6.04 \\
        \textbf{LIMBO230810} & 332.5~$\pm$~2.2 & 201~$\pm$~33 & 5.91 & 5.63 \\ \hline
    \end{tabular*}
    \label{tab:frb_props}
\end{table*}

Figure \ref{fig:misses} (\emph{top}) shows example spectra representative of the 7 events classified as RFI-induced false-positives.
In contrast to those in Figure \ref{fig:hits}, the RFI-induced events show broad or variable time and frequency structures, making them easy to rule out as FRBs.

Sometimes, genuine low-SNR bursts can be partially or fully obscured in files with RFI. 
As described in Section \ref{sec:analysis}, spectra are flagged, dedispersed, and inpainted across flagged channels to stabilize downstream statistics. 
In RFI-heavy intervals, this procedure introduces a sharp visual boundary between "clean" (inpainted) and "dirty" (unflagged but contaminated) regions in the dynamic spectrum. 
After dedispersion, this boundary shifts, producing the negatively sloped feature seen clearly in RFI230524 (Figure \ref{fig:misses}).

If an FRB overlaps such boundaries, only a subset of channels may retain astrophysical signal, while others are masked and replaced with Gaussian noise. 
This yields candidate FRBs that appear to occupy only part of LIMBO’s passband.
We observe this behaviour in 3 of the 24 files (Figure \ref{fig:misses} (\emph{bottom})). 
Although these events exhibit pulses with DMs consistent with \sgr, heavy contamination from RFI and inpainting prevents us from assessing their FRB candidacy with confidence. 
At present, LIMBO’s search pipeline does not attempt a dedicated recovery of such "hidden” FRBs, and they are excluded from our secure sample.
% XXX ARP mention how this is treated in our rate calculations?

\subsection{\sgr Rates} \label{sec:rates}

Our goal is to estimate the annual event rate from \sgr as a function of fluence, $R(\geq\mathcal{F})$.
It is possible that our sample contains false-positives.
The false-positive expectation number can be determined by integrating the tail of the cumulative distribution shown in Figure \ref{fig:EVD stats}. 
However, this distribution is a function of Z-score and systematics in our Z-scores make it difficult to translate expectations from that distribution into fluence space.

To address this, we average fluence errors for each burst to establish an approximate relationship between an $n\sigma$ event and its corresponding fluence. 
This yields $1\sigma\approx13~\mathrm{Jy\cdot ms}$. LIMBO230810 is excluded from this average because its noise statistics are noticeably different from the rest in the sample. 
Using this relation, we set the first of our fluence bins at the $5\sigma$ level ($\mathcal{F}=65~\mathrm{Jy\cdot ms}$) and logarithmically space the subsequent bins to reduce biases from small-number statistics. 

The expected number of false-positives in a given fluence bin is obtained by integrating the Gaussian tail, 
$$N_\mathrm{false}(b_{i}, b_{i+1}) = \frac{N_\mathrm{trials}}{2} \left[ \mathrm{erf}\left(\frac{b_{i+1}}{\sqrt{2}\sigma} \right) - \mathrm{erf}\left( \frac{b_i}{\sqrt{2}\sigma}\right ) \right ]$$
where $b_i$ and $b_{i+1}$ are the lower and upper fluence edges of the bin, respectively, and $\sigma=13~\mathrm{Jy\cdot ms}$.

Applying this formulation, 11 bursts fall in the $\mathcal{F} \geq~65~\mathrm{Jy\cdot ms}$ bin and 1 burst in the $\mathcal{F}\geq~130~\mathrm{Jy\cdot ms}$ bin. 
At most 1 spurious event is expected in the first bin and none in the second. 
These values are subtracted from the observed counts before rate estimation.

Sampling from a Poisson distribution then yields $R(\geq65~\mathrm{Jy \cdot ms}) = 112.3^{+81.3}_{-54.5}~\mathrm{yr}^{-1}$ and $R(\geq 130~\mathrm{Jy \cdot ms}) = 17.7^{+40.8}_{-15.1}~\mathrm{yr}^{-1}$, with uncertainties corresponding to 95\% confidence intervals.
The corresponding isotropic-equivalent burst energies for each of these fluence thresholds are $E_\mathrm{iso} = 8.8\times10^{29}~\rm erg$ and $1.8\times10^{30}~\rm erg$, respectively, assuming a distance of $9.5~\rm kpc$ to \sgr. 
These energies are comparable to the low-energy bursts detected by \cite{Kirsten2020}.
In contrast, the $E_\mathrm{iso}\sim10^{27}~\mathrm{erg}$ burst reported by \citet{FAST_SGR2020} represents a substantially lower energy event, whereas the exceptionally bright, FRB-like burst detected by STARE2 and CHIME had $E_\mathrm{iso}\sim10^{34}-10^{35}~\mathrm{erg}$ \citep{Bochenek2020, CHIME_SGR2020}.
Our LIMBO bursts remain several orders of magnitude less energetic than typical extragalactic FRBS, which generally exhibit isotropic-equivalent radio energies of $E_\mathrm{iso}\sim10^{35}-10^{42}~\mathrm{erg}$ \citep{CHIME_SGR2020}.
As such, it remains unclear whether these low-energy bursts produced by Galactic magnetars like \sgr are generated by the same physical mechanism as extragalactic FRBs, or whether they represent a distinct population of magnetar radio activity \citep{Kirsten2020}.

\begin{figure*}
    \centering
    \includegraphics[width=0.8\linewidth]{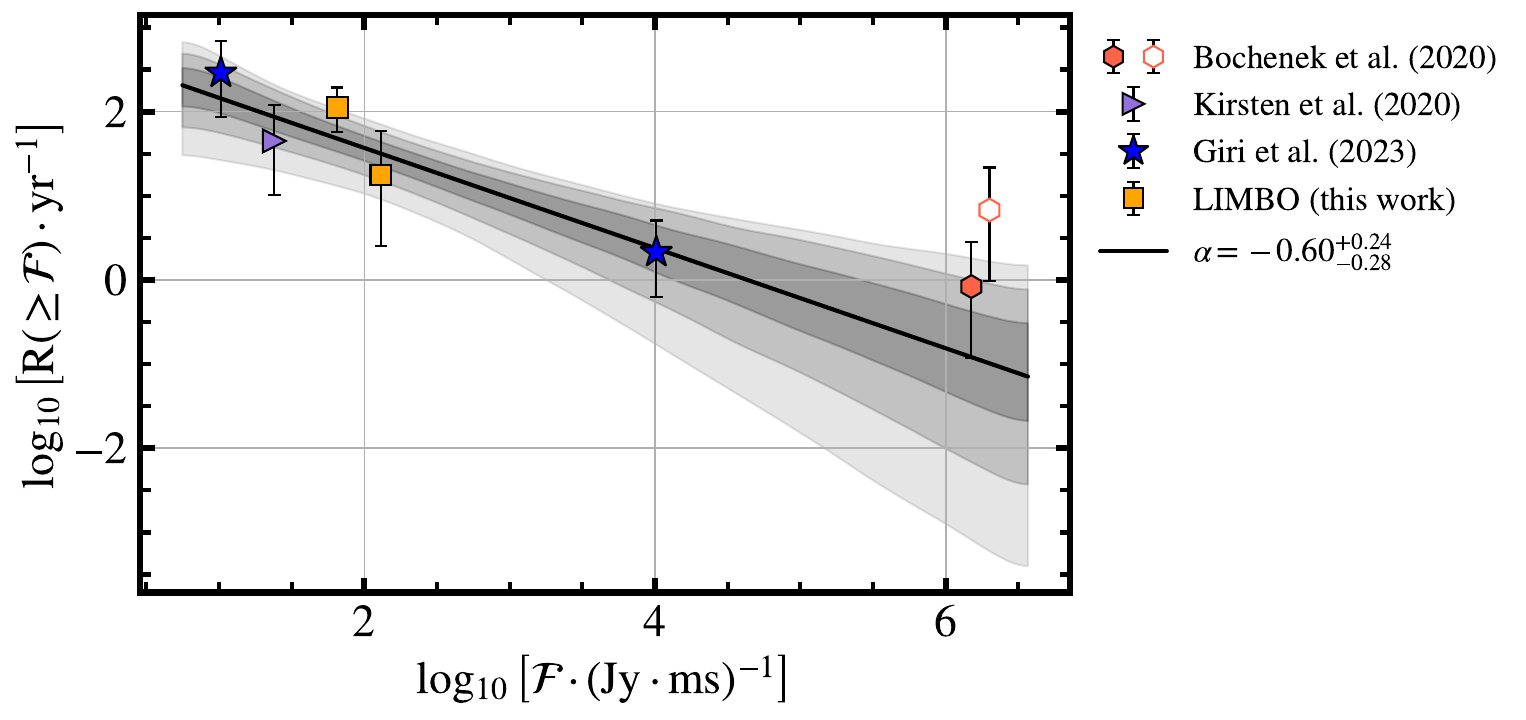}
    \caption{Cumulative burst rate as a function of fluence for FRBs associated with SGR~1935+2154. Rates derived from \citet{Kirsten2020} and \citet{Giri2023} are shown alongside the LIMBO results from this work, with uncertainties derived from their respective Poisson likelihoods. Two representations of the STARE2 \citep{Bochenek2020} $1.5~\mathrm{Jy\cdot ms}$ burst are shown: a population-averaged rate (solid pink), which accounts for look-elsewhere-effect corrections and marginalizes the uncertainty on the number of unknown Galactic magnetars, and a source-conditioned rate (white marker with pink outline), accounting only for the effective observing time in which SGR~1935+2154 was visible to STARE2 and Poisson uncertainty alone. The population-averaged marker is offset slightly in fluence for visual clarity. All rate uncertainties correspond to a 95\% confidence interval. The solid black line shows the line of best fit for the power-law model to the cumulative rate-fluence distribution, with a slope of $\alpha=-0.60^{+0.24}_{-0.28}$ (95\% CI). The fit excludes the source-conditioned STARE2 rate. The shaded regions, from dark to light gray, correspond to the 68\%, 95\%, and 99.7\% highest-density intervals of the model posterior, respectively.}
    \label{fig:rates vs fluence}
\end{figure*}

To further facilitate a comparison of our work, Figure \ref{fig:rates vs fluence} compares the LIMBO rates to those inferred from other well-documented \sgr FRBs-like bursts \citep{Bochenek2020, Kirsten2020, Giri2023}. 
As was done for LIMBO, uncertainties in the rates for \citet{Kirsten2020} and \citet{Giri2023} are derived by sampling their respective Poisson likelihoods. 
A power-law model is fit to the cumulative rate-fluence distribution using the Markov chain Monte Carlo (MCMC) sampler \texttt{emcee} \citep{emcee}, yielding a slope of $\alpha=-0.60^{+0.24}_{-0.28}$, where uncertainties correspond to the 95\% highest-density interval. 

We show two representations of the $1.5~\mathrm{MJy \cdot ms}$ burst detected by STARE2 \citep{Bochenek2020}. 
The first corresponds to a population-averaged rate (solid pink), which accounts for the look-elsewhere effect arising from STARE2's all-sky monitoring of multiple Galactic magnetars over an extended observing time. 
In this case, the effective observing time is obtained by summing the visibility-weighted observing times over the entire known Galactic magnetar population. 
The inferred rate is further marginalized over uncertainty in the number of currently unknown Galactic magnetars by weighting STARE2's Poisson likelihood (modeled as a Gamma distribution) with an exponential prior with a scale of $\sigma=1$ on the number of unseen magnetars. 
We find that variations in the scale ($\sigma=0.01-10$) alter the inferred power-law slope by less than 4\%, and therefore do not change our conclusions.

For comparison, we also show a source-conditioned rate using only the effective observing time during which \sgr itself was visible to STARE2 (white marker with pink outline).
This point reflects Poisson uncertainly alone without additional population-averaging or look-elsewhere corrections. 
Although the STARE2 burst is associated with \sgr, the detection resulted from an untargeted, all-sky experiment. 
The inferred rate therefore reflects a draw from the Galactic magnetar population rather than a source-conditioned rate. 
As a result, only the population-averaged rate and corresponding likelihood is used in the rate-fluence power-law fit.

As discussed in Section \ref{sec:sgr} and Appendix \ref{appx:rfis}, three events with $Z_\mathrm{max}\geq 5.6$ exhibit FRB-like features but are heavily contaminated by RFI. If we treat these events as candidate FRBs, the inferred rate for \sgr shows a modest increase to $R(\geq65~\mathrm{Jy \cdot ms}) = 143.8^{+90.1}_{-63.3}~\mathrm{yr}^{-1}$. There is no change in $R(\geq130~\mathrm{Jy \cdot ms})$. Moreover, the consideration of these "hidden" bursts changes the inferred power-law slope, $\alpha$, by less than $5\%$.

\section{Conclusion} \label{sec:conclusions}

The Long-Integration Magnetar Burst Observatory (LIMBO) is a real-time radio transient detection pipeline specifically designed to conduct directed, long-term observations of Galactic magnetars to detect and characterize fast radio bursts (FRBs).
LIMBO is currently deployed on the UC Berkeley Leuschner Radio Observatory's \leuschsize radio dish, equipped with a dual-polarization feed horn and operating with a 125~MHz passband centred at 1460~MHz.
A real-time processing pipeline searches for dispersed transients in the combined polarizations using a modified Fast Dispersion Measure Transform (FDMT) algorithm for coherent dedispersion of bursts, and triggers a dump of voltage data to drive when a candidate FRB is detected.
Using a combination of on-sky calibrations, synthetic signal-injection and recovery tests, and the successful detection of giant pulses from the Crab Pulsar, we validate LIMBO's sensitivity to radio transients with fluences $\geq 43~\mathrm{Jy \cdot ms}$.
LIMBO's sensitivity to faint Galactic FRBs and capacity to carry-out $\sim10^4$~hours of follow-up observations allow LIMBO to explore an otherwise untouched parameter space in FRB detections (Figure \ref{fig:frb_followup}).

Between May and August 2023, LIMBO conducted 833~hrs of follow-up of Galactic magnetar \sgr. 
We identify 12 low-energy bursts tentatively classified as candidate FRBs, yielding inferred rates of $R(\geq65~\mathrm{Jy \cdot ms}) = 112.3^{+81.3}_{-54.5}~\mathrm{yr}^{-1}$ and $R(\geq 130~\mathrm{Jy \cdot ms}) = 17.7^{+40.8}_{-15.1}~\mathrm{yr}^{-1}$.
These bursts are compared with other detected FRB-like bursts from \sgr, and a cumulative rate-fluence power-law slope of $\alpha=-0.60^{+0.24}_{-0.28}$ is inferred (Figure \ref{fig:rates vs fluence}).

Beyond expanding the census of Galactic FRBs, LIMBO provides a critical step toward connecting burst phenomenology with the underlying emission physics. 
Long-duration, directed monitoring of a single, well-characterized source enables constraints on burst clustering, fluence distributions, polarization properties, and temporal correlations with high-energy activity-observables that are difficult to access in wide-field, discovery-driven surveys. 
The ability to trigger voltage dumps further allows for high-time-resolution studies of burst substructure, coherence, and polarization evolution, all of which encode information about the emission region and mechanism.

In this sense, LIMBO complements large population surveys by shifting the emphasis from detection rates alone to physically motivated characterization of individual sources.
As such, LIMBO helps bridge the gap between cosmological FRB statistics and source-resolved studies, providing a pathway toward distinguishing between competing emission models and clarifying the role of magnetars as FRB progenitors.

Future work with LIMBO will focus on detailed analysis of voltage data for the candidates presented in this work, alongside continued searches for and characterization of new candidates in ongoing follow-up observations of \sgr and other Galactic magnetars.

%%%%%%%%%%%%%%%%%%%%%%%%%%%%%%%%%%%%%%%%%%%%%%%%%%
\section*{Acknowledgments}
D.M. acknowledges Matt Dexter and Christian Hellum Bye for early guidance and valuable discussions, particularly regarding the design of the LIMBO instrument and development of the transient detection pipeline. D.M. further thanks Padma Venkatraman for valuable discussion and advice regarding statistical analysis.

%%%%%%%%%%%%%%%%%%%%%%%%%%%%%%%%%%%%%%%%%%%%%%%%%%
\section*{Conflicts of Interest}
Authors declare no conflict of interest.

%%%%%%%%%%%%%%%%%%%%%%%%%%%%%%%%%%%%%%%%%%%%%%%%%%
\section*{Data Availability}
Calibrating data for \sgr statistics as well as data containing the 24 events analysed for candidate FRBs can be found at \url{https://doi.org/10.5281/zenodo.18870723}. The LIMBO processing pipeline and telescope control software can be found at \url{https://github.com/AaronParsons/limbo}. Data recording backends used for LIMBO are found at \url{https://github.com/liuweiseu/limbo_recorder/tree/auto_files}, while control of the LIMBO SNAP board is found at \url{https://github.com/liuweiseu/limbo_scripts}.

%%%%%%%%%%%%%%%%%%%% REFERENCES %%%%%%%%%%%%%%%%%%

% The best way to enter references is to use BibTeX:

\bibliographystyle{rasti}
\bibliography{references} % if your bibtex file is called example.bib

@ARTICLE{Jia2026,
       author = {{Jia}, X.~D. and {Gao}, D.~H. and {Chen}, J.~H. and {Wu}, Q. and {Yi}, S.~X. and {Wang}, F.~Y.},
        title = "{Cosmological Evolution of Fast Radio Bursts and Its Rapid Decline Relative to Star Formation Rate}",
      journal = {\apj},
         year = 2026,
        month = jun,
       volume = {1003},
       number = {2},
          eid = {179},
        pages = {179},
          doi = {10.3847/1538-4357/ae665c},
archivePrefix = {arXiv},
       eprint = {2604.26574},
 primaryClass = {astro-ph.HE},
       adsurl = {https://ui.adsabs.harvard.edu/abs/2026ApJ..1003..179J}
}

@ARTICLE{Lei2025,
       author = {{Lei}, Qing-Zhen and {Wang}, Xin-Zhe and {Deng}, Can-Min},
        title = "{A Comprehensive Study of the Energy and Redshift Distributions of the Fast Radio Burst Population Based on the First CHIME/FRB Catalog}",
      journal = {\apj},
         year = 2025,
        month = sep,
       volume = {990},
       number = {2},
          eid = {175},
        pages = {175},
          doi = {10.3847/1538-4357/adf646},
archivePrefix = {arXiv},
       eprint = {2507.23122},
 primaryClass = {astro-ph.HE},
       adsurl = {https://ui.adsabs.harvard.edu/abs/2025ApJ...990..175L}
}

@ARTICLE{Eftekhari2017,
       author = {{Eftekhari}, T. and {Berger}, E.},
        title = "{Associating Fast Radio Bursts with Their Host Galaxies}",
      journal = {\apj},
         year = 2017,
        month = nov,
       volume = {849},
       number = {2},
          eid = {162},
        pages = {162},
          doi = {10.3847/1538-4357/aa90b9},
archivePrefix = {arXiv},
       eprint = {1705.02998},
 primaryClass = {astro-ph.HE},
       adsurl = {https://ui.adsabs.harvard.edu/abs/2017ApJ...849..162E}
}

@ARTICLE{Gordon2025,
       author = {{Gordon}, Alexa C. and {Fong}, Wen-fai and {Deller}, Adam T. and {Marnoch}, Lachlan and {Lim}, Sungsoon and {Peng}, Eric W. and {Bannister}, Keith W. and {Bera}, Apurba and {Bhat}, N.~D.~R. and {Dial}, Tyson and {Dong}, Yuxin and {Eftekhari}, Tarraneh and {Glowacki}, Marcin and {Gourdji}, Kelly and {Gupta}, Vivek and {Jahns-Schindler}, Joscha N. and {Jaini}, Akhil and {Kilpatrick}, Charles D. and {Liu}, Chang and {Prochaska}, J. Xavier and {Ryder}, Stuart D. and {Shannon}, Ryan M. and {Simha}, Sunil and {Tejos}, Nicolas and {Wang}, Yuanming and {Wang}, Ziteng},
        title = "{Mapping the Spatial Distribution of Fast Radio Bursts within their Host Galaxies}",
      journal = {\apj},
         year = 2025,
        month = nov,
       volume = {993},
       number = {1},
          eid = {119},
        pages = {119},
          doi = {10.3847/1538-4357/ae0298},
archivePrefix = {arXiv},
       eprint = {2506.06453},
 primaryClass = {astro-ph.GA},
       adsurl = {https://ui.adsabs.harvard.edu/abs/2025ApJ...993..119G}
}

@ARTICLE{Macquart2018,
       author = {{Macquart}, J.-P. and {Ekers}, RD},
        title = "{FRB event rate counts - II. Fluence, redshift, and dispersion measure distributions}",
      journal = {\mnras},
         year = 2018,
        month = nov,
       volume = {480},
       number = {3},
        pages = {4211-4230},
          doi = {10.1093/mnras/sty2083},
archivePrefix = {arXiv},
       eprint = {1808.00908},
 primaryClass = {astro-ph.HE},
       adsurl = {https://ui.adsabs.harvard.edu/abs/2018MNRAS.480.4211M}
}

@ARTICLE{Zhang2021,
       author = {{Zhang}, Rachel C. and {Zhang}, Bing and {Li}, Ye and {Lorimer}, Duncan R.},
        title = "{On the energy and redshift distributions of fast radio bursts}",
      journal = {\mnras},
         year = 2021,
        month = feb,
       volume = {501},
       number = {1},
        pages = {157-167},
          doi = {10.1093/mnras/staa3537},
archivePrefix = {arXiv},
       eprint = {2011.06151},
 primaryClass = {astro-ph.HE},
       adsurl = {https://ui.adsabs.harvard.edu/abs/2021MNRAS.501..157Z}
}

@ARTICLE{Connor2020,
       author = {{Connor}, L. and {van Leeuwen}, J. and {Oostrum}, L.~C. and {Petroff}, E. and {Maan}, Y. and {Adams}, E.~A.~K. and {Attema}, J.~J. and {Bast}, J.~E. and {Boersma}, O.~M. and {D{\'e}nes}, H. and {Gardenier}, D.~W. and {Hargreaves}, J.~E. and {Kooistra}, E. and {Pastor-Marazuela}, I. and {Schulz}, R. and {Sclocco}, A. and {Smits}, R. and {Straal}, S.~M. and {van der Schuur}, D. and {Vohl}, D. and {Adebahr}, B. and {de Blok}, W.~J.~G. and {van Cappellen}, W.~A. and {Coolen}, A.~H.~W.~M. and {Damstra}, S. and {van Diepen}, G.~N.~J. and {Frank}, B.~S. and {Hess}, K.~M. and {Hut}, B. and {Kutkin}, A. and {Loose}, G. Marcel and {Lucero}, D.~M. and {Mika}, {\'A}. and {Moss}, V.~A. and {Mulder}, H. and {Oosterloo}, T.~A. and {Ruiter}, M. and {Vedantham}, H. and {Vermaas}, N.~J. and {Wijnholds}, S.~J. and {Ziemke}, J.},
        title = "{A bright, high rotation-measure FRB that skewers the M33 halo}",
      journal = {\mnras},
         year = 2020,
        month = dec,
       volume = {499},
       number = {4},
        pages = {4716-4724},
          doi = {10.1093/mnras/staa3009},
archivePrefix = {arXiv},
       eprint = {2002.01399},
 primaryClass = {astro-ph.HE},
       adsurl = {https://ui.adsabs.harvard.edu/abs/2020MNRAS.499.4716C}
}

@ARTICLE{Farah2019,
       author = {{Farah}, W. and {Flynn}, C. and {Bailes}, M. and {Jameson}, A. and {Bateman}, T. and {Campbell-Wilson}, D. and {Day}, C.~K. and {Deller}, A.~T. and {Green}, A.~J. and {Gupta}, V. and {Hunstead}, R. and {Lower}, M.~E. and {Os{\l}owski}, S. and {Parthasarathy}, A. and {Price}, D.~C. and {Ravi}, V. and {Shannon}, R.~M. and {Sutherland}, A. and {Temby}, D. and {Krishnan}, V. Venkatraman and {Caleb}, M. and {Chang}, S.-W. and {Cruces}, M. and {Roy}, J. and {Morello}, V. and {Onken}, C.~A. and {Stappers}, B.~W. and {Webb}, S. and {Wolf}, C.},
        title = "{Five new real-time detections of fast radio bursts with UTMOST}",
      journal = {\mnras},
         year = 2019,
        month = sep,
       volume = {488},
       number = {3},
        pages = {2989-3002},
          doi = {10.1093/mnras/stz1748},
archivePrefix = {arXiv},
       eprint = {1905.02293},
 primaryClass = {astro-ph.HE},
       adsurl = {https://ui.adsabs.harvard.edu/abs/2019MNRAS.488.2989F}
}

@ARTICLE{Masui2015,
       author = {{Masui}, Kiyoshi and {Lin}, Hsiu-Hsien and {Sievers}, Jonathan and {Anderson}, Christopher J. and {Chang}, Tzu-Ching and {Chen}, Xuelei and {Ganguly}, Apratim and {Jarvis}, Miranda and {Kuo}, Cheng-Yu and {Li}, Yi-Chao and {Liao}, Yu-Wei and {McLaughlin}, Maura and {Pen}, Ue-Li and {Peterson}, Jeffrey B. and {Roman}, Alexander and {Timbie}, Peter T. and {Voytek}, Tabitha and {Yadav}, Jaswant K.},
        title = "{Dense magnetized plasma associated with a fast radio burst}",
      journal = {\nat},
         year = 2015,
        month = dec,
       volume = {528},
       number = {7583},
        pages = {523-525},
          doi = {10.1038/nature15769},
archivePrefix = {arXiv},
       eprint = {1512.00529},
 primaryClass = {astro-ph.HE},
       adsurl = {https://ui.adsabs.harvard.edu/abs/2015Natur.528..523M}
}

@ARTICLE{Majid2021,
       author = {{Majid}, Walid A. and {Pearlman}, Aaron B. and {Prince}, Thomas A. and {Wharton}, Robert S. and {Naudet}, Charles J. and {Bansal}, Karishma and {Connor}, Liam and {Bhardwaj}, Mohit and {Tendulkar}, Shriharsh P.},
        title = "{A Bright Fast Radio Burst from FRB 20200120E with Sub-100 Nanosecond Structure}",
      journal = {\apjl},
         year = 2021,
        month = sep,
       volume = {919},
       number = {1},
          eid = {L6},
        pages = {L6},
          doi = {10.3847/2041-8213/ac1921},
archivePrefix = {arXiv},
       eprint = {2105.10987},
 primaryClass = {astro-ph.HE},
       adsurl = {https://ui.adsabs.harvard.edu/abs/2021ApJ...919L...6M}
}

@ARTICLE{Zhu2020,
       author = {{Zhu}, Weiwei and {Li}, Di and {Luo}, Rui and {Miao}, Chenchen and {Zhang}, Bing and {Spitler}, Laura and {Lorimer}, Duncan and {Kramer}, Michael and {Champion}, David and {Yue}, Youling and {Cameron}, Andrew and {Cruces}, Marilyn and {Duan}, Ran and {Feng}, Yi and {Han}, Jun and {Hobbs}, George and {Niu}, Chenhui and {Niu}, Jiarui and {Pan}, Zhichen and {Qian}, Lei and {Shi}, Dai and {Tang}, Ningyu and {Wang}, Pei and {Wang}, Hongfeng and {Yuan}, Mao and {Zhang}, Lei and {Zhang}, Xinxin and {Cao}, Shuyun and {Feng}, Li and {Gan}, Hengqian and {Gao}, Long and {Gu}, Xuedong and {Guo}, Minglei and {Hao}, Qiaoli and {Huang}, Lin and {Huang}, Menglin and {Jiang}, Peng and {Jin}, Chengjin and {Li}, Hui and {Li}, Qi and {Li}, Qisheng and {Liu}, Hongfei and {Pan}, Gaofeng and {Peng}, Bo and {Qian}, Hui and {Shi}, Xiangwei and {Song}, Jinyuo and {Song}, Liqiang and {Sun}, Caihong and {Sun}, Jinghai and {Wang}, Hong and {Wang}, Qiming and {Wang}, Yi and {Xie}, Xiaoyao and {Yan}, Jun and {Yang}, Li and {Yang}, Shimo and {Yao}, Rui and {Yu}, Dongjun and {Yu}, Jinglong and {Zhang}, Chengmin and {Zhang}, Haiyan and {Zhang}, Shuxin and {Zheng}, Xiaonian and {Zhou}, Aiying and {Zhu}, Boqin and {Zhu}, Lichun and {Zhu}, Ming and {Zhu}, Wenbai and {Zhu}, Yan},
        title = "{A Fast Radio Burst Discovered in FAST Drift Scan Survey}",
      journal = {\apjl},
         year = 2020,
        month = may,
       volume = {895},
       number = {1},
          eid = {L6},
        pages = {L6},
          doi = {10.3847/2041-8213/ab8e46},
archivePrefix = {arXiv},
       eprint = {2004.14029},
 primaryClass = {astro-ph.HE},
       adsurl = {https://ui.adsabs.harvard.edu/abs/2020ApJ...895L...6Z}
}

@ARTICLE{Niu2021,
       author = {{Niu}, Chen-Hui and {Li}, Di and {Luo}, Rui and {Wang}, Wei-Yang and {Yao}, Jumei and {Zhang}, Bing and {Zhu}, Wei-Wei and {Wang}, Pei and {Ye}, Haoyang and {Zhang}, Yong-Kun and {Niu}, Jia-rui and {Tang}, Ning-yu and {Duan}, Ran and {Krco}, Marko and {Dai}, Shi and {Feng}, Yi and {Miao}, Chenchen and {Pan}, Zhichen and {Qian}, Lei and {Xue}, Mengyao and {Yuan}, Mao and {Yue}, Youling and {Zhang}, Lei and {Zhang}, Xinxin},
        title = "{CRAFTS for Fast Radio Bursts: Extending the Dispersion-Fluence Relation with New FRBs Detected by FAST}",
      journal = {\apjl},
         year = 2021,
        month = mar,
       volume = {909},
       number = {1},
          eid = {L8},
        pages = {L8},
          doi = {10.3847/2041-8213/abe7f0},
archivePrefix = {arXiv},
       eprint = {2102.10546},
 primaryClass = {astro-ph.HE},
       adsurl = {https://ui.adsabs.harvard.edu/abs/2021ApJ...909L...8N}
}

@ARTICLE{Luo2020,
       author = {{Luo}, R. and {Wang}, B.~J. and {Men}, Y.~P. and {Zhang}, C.~F. and {Jiang}, J.~C. and {Xu}, H. and {Wang}, W.~Y. and {Lee}, K.~J. and {Han}, J.~L. and {Zhang}, B. and {Caballero}, R.~N. and {Chen}, M.~Z. and {Chen}, X.~L. and {Gan}, H.~Q. and {Guo}, Y.~J. and {Hao}, L.~F. and {Huang}, Y.~X. and {Jiang}, P. and {Li}, H. and {Li}, J. and {Li}, Z.~X. and {Luo}, J.~T. and {Pan}, J. and {Pei}, X. and {Qian}, L. and {Sun}, J.~H. and {Wang}, M. and {Wang}, N. and {Wen}, Z.~G. and {Xu}, R.~X. and {Xu}, Y.~H. and {Yan}, J. and {Yan}, W.~M. and {Yu}, D.~J. and {Yuan}, J.~P. and {Zhang}, S.~B. and {Zhu}, Y.},
        title = "{Diverse polarization angle swings from a repeating fast radio burst source}",
      journal = {\nat},
         year = 2020,
        month = oct,
       volume = {586},
       number = {7831},
        pages = {693-696},
          doi = {10.1038/s41586-020-2827-2},
archivePrefix = {arXiv},
       eprint = {2011.00171},
 primaryClass = {astro-ph.HE},
       adsurl = {https://ui.adsabs.harvard.edu/abs/2020Natur.586..693L}
}

@ARTICLE{Hashimoto2022,
       author = {{Hashimoto}, Tetsuya and {Goto}, Tomotsugu and {Chen}, Bo Han and {Ho}, Simon C.-C. and {Hsiao}, Tiger Y.-Y. and {Wong}, Yi Hang Valerie and {On}, Alvina Y.~L. and {Kim}, Seong Jin and {Kilerci-Eser}, Ece and {Huang}, Kai-Chun and {Santos}, Daryl Joe D. and {Yamasaki}, Shotaro},
        title = "{Energy functions of fast radio bursts derived from the first CHIME/FRB catalogue}",
      journal = {\mnras},
         year = 2022,
        month = apr,
       volume = {511},
       number = {2},
        pages = {1961-1976},
          doi = {10.1093/mnras/stac065},
archivePrefix = {arXiv},
       eprint = {2201.03574},
 primaryClass = {astro-ph.HE},
       adsurl = {https://ui.adsabs.harvard.edu/abs/2022MNRAS.511.1961H}
}

@ARTICLE{Wharton2025,
       author = {{Wharton}, Robert S. and {Majid}, Walid A. and {Sherman}, Myles B. and {Connor}, Liam and {Kocz}, Jonathon and {Lu}, Wenbin and {Pearlman}, Aaron B. and {Prince}, Thomas A. and {Ravi}, Vikram},
        title = "{High-frequency Fast Radio Burst Search of Nearby Star-forming Galaxies M77 and M82}",
      journal = {\apj},
         year = 2025,
        month = may,
       volume = {984},
       number = {2},
          eid = {119},
        pages = {119},
          doi = {10.3847/1538-4357/adc303},
       adsurl = {https://ui.adsabs.harvard.edu/abs/2025ApJ...984..119W}
}

@ARTICLE{Geminardi2025,
       author = {{Geminardi}, A. and {Esposito}, P. and {Bernardi}, G. and {Pilia}, M. and {Pelliciari}, D. and {Naldi}, G. and {Dallacasa}, D. and {Turolla}, R. and {Stella}, L. and {Perini}, F. and {Verrecchia}, F. and {Casentini}, C. and {Trudu}, M. and {Lulli}, R. and {Maccaferri}, A. and {Magro}, A. and {Mattana}, A. and {Bianchi}, G. and {Pupillo}, G. and {Bortolotti}, C. and {Tavani}, M. and {Roma}, M. and {Schiaffino}, M. and {Setti}, G.},
        title = "{The Northern Cross Fast Radio Burst project: V. Search for transient radio emission from Galactic magnetars}",
      journal = {\aap},
         year = 2025,
        month = aug,
       volume = {700},
          eid = {A19},
        pages = {A19},
          doi = {10.1051/0004-6361/202554386},
archivePrefix = {arXiv},
       eprint = {2505.24049},
 primaryClass = {astro-ph.HE},
       adsurl = {https://ui.adsabs.harvard.edu/abs/2025A&A...700A..19G}
}

@ARTICLE{Pelliciari2023,
       author = {{Pelliciari}, D. and {Bernardi}, G. and {Pilia}, M. and {Naldi}, G. and {Pupillo}, G. and {Trudu}, M. and {Addis}, A. and {Bianchi}, G. and {Bortolotti}, C. and {Dallacasa}, D. and {Lulli}, R. and {Maccaferri}, A. and {Magro}, A. and {Mattana}, A. and {Perini}, F. and {Roma}, M. and {Schiaffino}, M. and {Setti}, G. and {Tavani}, M. and {Verrecchia}, F. and {Casentini}, C.},
        title = "{The Northern Cross Fast Radio Burst project. III. The FRB-magnetar connection in a sample of nearby galaxies}",
      journal = {\aap},
         year = 2023,
        month = jun,
       volume = {674},
          eid = {A223},
        pages = {A223},
          doi = {10.1051/0004-6361/202346307},
archivePrefix = {arXiv},
       eprint = {2304.11179},
 primaryClass = {astro-ph.HE},
       adsurl = {https://ui.adsabs.harvard.edu/abs/2023A&A...674A.223P}
}

@ARTICLE{CordesWasserman2016,
       author = {{Cordes}, J.~M. and {Wasserman}, Ira},
        title = "{Supergiant pulses from extragalactic neutron stars}",
      journal = {\mnras},
         year = 2016,
        month = mar,
       volume = {457},
       number = {1},
        pages = {232-257},
          doi = {10.1093/mnras/stv2948},
archivePrefix = {arXiv},
       eprint = {1501.00753},
 primaryClass = {astro-ph.HE},
       adsurl = {https://ui.adsabs.harvard.edu/abs/2016MNRAS.457..232C}
}

@ARTICLE{Mingarelli2015,
       author = {{Mingarelli}, Chiara M.~F. and {Levin}, Janna and {Lazio}, T. Joseph W.},
        title = "{Fast Radio Bursts and Radio Transients from Black Hole Batteries}",
      journal = {\apjl},
         year = 2015,
        month = dec,
       volume = {814},
       number = {2},
          eid = {L20},
        pages = {L20},
          doi = {10.1088/2041-8205/814/2/L20},
archivePrefix = {arXiv},
       eprint = {1511.02870},
 primaryClass = {astro-ph.HE},
       adsurl = {https://ui.adsabs.harvard.edu/abs/2015ApJ...814L..20M}
}

@ARTICLE{FalckeRezzolla2014,
       author = {{Falcke}, Heino and {Rezzolla}, Luciano},
        title = "{Fast radio bursts: the last sign of supramassive neutron stars}",
      journal = {\aap},
         year = 2014,
        month = feb,
       volume = {562},
          eid = {A137},
        pages = {A137},
          doi = {10.1051/0004-6361/201321996},
archivePrefix = {arXiv},
       eprint = {1307.1409},
 primaryClass = {astro-ph.HE},
       adsurl = {https://ui.adsabs.harvard.edu/abs/2014A&A...562A.137F}
}

@ARTICLE{Totani2013,
       author = {{Totani}, Tomonori},
        title = "{Cosmological Fast Radio Bursts from Binary Neutron Star Mergers}",
      journal = {\pasj},
         year = 2013,
        month = oct,
       volume = {65},
       number = {5},
          eid = {L12},
        pages = {L12},
          doi = {10.1093/pasj/65.5.L12},
archivePrefix = {arXiv},
       eprint = {1307.4985},
 primaryClass = {astro-ph.HE},
       adsurl = {https://ui.adsabs.harvard.edu/abs/2013PASJ...65L..12T}
}

@ARTICLE{Beloborodov2017,
       author = {{Beloborodov}, Andrei M.},
        title = "{A Flaring Magnetar in FRB 121102?}",
      journal = {\apjl},
         year = 2017,
        month = jul,
       volume = {843},
       number = {2},
          eid = {L26},
        pages = {L26},
          doi = {10.3847/2041-8213/aa78f3},
archivePrefix = {arXiv},
       eprint = {1702.08644},
 primaryClass = {astro-ph.HE},
       adsurl = {https://ui.adsabs.harvard.edu/abs/2017ApJ...843L..26B}
}

@ARTICLE{Murase2016,
       author = {{Murase}, Kohta and {Kashiyama}, Kazumi and {M{\'e}sz{\'a}ros}, Peter},
        title = "{A burst in a wind bubble and the impact on baryonic ejecta: high-energy gamma-ray flashes and afterglows from fast radio bursts and pulsar-driven supernova remnants}",
      journal = {\mnras},
         year = 2016,
        month = sep,
       volume = {461},
       number = {2},
        pages = {1498-1511},
          doi = {10.1093/mnras/stw1328},
archivePrefix = {arXiv},
       eprint = {1603.08875},
 primaryClass = {astro-ph.HE},
       adsurl = {https://ui.adsabs.harvard.edu/abs/2016MNRAS.461.1498M}
}

@ARTICLE{Lyubarsky2014,
       author = {{Lyubarsky}, Yu.},
        title = "{A model for fast extragalactic radio bursts.}",
      journal = {\mnras},
         year = 2014,
        month = jul,
       volume = {442},
        pages = {L9-L13},
          doi = {10.1093/mnrasl/slu046},
archivePrefix = {arXiv},
       eprint = {1401.6674},
 primaryClass = {astro-ph.HE},
       adsurl = {https://ui.adsabs.harvard.edu/abs/2014MNRAS.442L...9L}
}

@INPROCEEDINGS{PopovPostnov2007,
       author = {{Popov}, Sergey B. and {Postnov}, K.~A.},
        title = "{Hyperflares of SGRs as an engine for millisecond extragalactic radio bursts}",
    booktitle = {Evolution of Cosmic Objects through their Physical Activity},
         year = 2010,
       editor = {{Harutyunian}, H.~A. and {Mickaelian}, A.~M. and {Terzian}, Y.},
        month = nov,
        pages = {129-132},
          doi = {10.48550/arXiv.0710.2006},
archivePrefix = {arXiv},
       eprint = {0710.2006},
 primaryClass = {astro-ph},
       adsurl = {https://ui.adsabs.harvard.edu/abs/2010vaoa.conf..129P}
}

@ARTICLE{PetroffHesselsLorimer2019,
       author = {{Petroff}, E. and {Hessels}, J.~W.~T. and {Lorimer}, D.~R.},
        title = "{Fast radio bursts}",
      journal = {\aapr},
         year = 2019,
        month = dec,
       volume = {27},
       number = {1},
          eid = {4},
        pages = {4},
          doi = {10.1007/s00159-019-0116-6},
archivePrefix = {arXiv},
       eprint = {1904.07947},
 primaryClass = {astro-ph.HE},
       adsurl = {https://ui.adsabs.harvard.edu/abs/2019A&ARv..27....4P}
}

@ARTICLE{Platts2019,
       author = {{Platts}, E. and {Weltman}, A. and {Walters}, A. and {Tendulkar}, S.~P. and {Gordin}, J.~E.~B. and {Kandhai}, S.},
        title = "{A living theory catalogue for fast radio bursts}",
      journal = {\physrep},
         year = 2019,
        month = aug,
       volume = {821},
        pages = {1-27},
          doi = {10.1016/j.physrep.2019.06.003},
archivePrefix = {arXiv},
       eprint = {1810.05836},
 primaryClass = {astro-ph.HE},
       adsurl = {https://ui.adsabs.harvard.edu/abs/2019PhR...821....1P}
}

@ARTICLE{Gordon2023,
       author = {{Gordon}, Alexa C. and {Fong}, Wen-fai and {Kilpatrick}, Charles D. and {Eftekhari}, Tarraneh and {Leja}, Joel and {Prochaska}, J. Xavier and {Nugent}, Anya E. and {Bhandari}, Shivani and {Blanchard}, Peter K. and {Caleb}, Manisha and {Day}, Cherie K. and {Deller}, Adam T. and {Dong}, Yuxin and {Glowacki}, Marcin and {Gourdji}, Kelly and {Mannings}, Alexandra G. and {Mahoney}, Elizabeth K. and {Marnoch}, Lachlan and {Miller}, Adam A. and {Paterson}, Kerry and {Rastinejad}, Jillian C. and {Ryder}, Stuart D. and {Sadler}, Elaine M. and {Scott}, Danica R. and {Sears}, Huei and {Shannon}, Ryan M. and {Simha}, Sunil and {Stappers}, Benjamin W. and {Tejos}, Nicolas},
        title = "{The Demographics, Stellar Populations, and Star Formation Histories of Fast Radio Burst Host Galaxies: Implications for the Progenitors}",
      journal = {\apj},
         year = 2023,
        month = sep,
       volume = {954},
       number = {1},
          eid = {80},
        pages = {80},
          doi = {10.3847/1538-4357/ace5aa},
archivePrefix = {arXiv},
       eprint = {2302.05465},
 primaryClass = {astro-ph.GA},
       adsurl = {https://ui.adsabs.harvard.edu/abs/2023ApJ...954...80G}
}

@ARTICLE{Shitov1999,
       author = {{Shitov}, Yu. P.},
        title = "{SGR 1900+14 = PSR J1907+0919}",
      journal = {\iaucirc},
         year = 1999,
        month = feb,
       volume = {7110},
        pages = {2},
       adsurl = {https://ui.adsabs.harvard.edu/abs/1999IAUC.7110....2S}
}

@ARTICLE{Younes2025,
       author = {{Younes}, George and {Lander}, Samuel K. and {Baring}, Matthew G. and {Bause}, Marlon L. and {Stewart}, Rachael and {Arzoumanian}, Zaven and {Thi}, Hoa Dinh and {Enoto}, Teruaki and {Gendreau}, Keith C. and {G{\"u}ver}, Tolga and {Harding}, Alice K. and {Ho}, Wynn C.~G. and {Hu}, Chin-Ping and {Van Kooten}, Alex and {Kouveliotou}, Chryssa and {Di Lalla}, Niccol{\`o} and {McEwen}, Alexander and {Negro}, Michela and {Ng}, Mason and {Palmer}, David M. and {Spitler}, Laura G. and {Wadiasingh}, Zorawar},
        title = "{Timing and Spectral Evolution of the Magnetar 1E 1841-045 in Outburst}",
      journal = {\apj},
         year = 2025,
        month = aug,
       volume = {989},
       number = {1},
          eid = {89},
        pages = {89},
          doi = {10.3847/1538-4357/ade716},
archivePrefix = {arXiv},
       eprint = {2502.20079},
 primaryClass = {astro-ph.HE},
       adsurl = {https://ui.adsabs.harvard.edu/abs/2025ApJ...989...89Y}
}

@ARTICLE{Hu2020,
       author = {{Hu}, Chin-Ping and {Begi{\c{c}}arslan}, Beste and {G{\"u}ver}, Tolga and {Enoto}, Teruaki and {Younes}, George and {Sakamoto}, Takanori and {Ray}, Paul S. and {Strohmayer}, Tod E. and {Guillot}, Sebastien and {Arzoumanian}, Zaven and {Palmer}, David M. and {Gendreau}, Keith C. and {Malacaria}, C. and {Wadiasingh}, Zorawar and {Jaisawal}, Gaurava K. and {Majid}, Walid A.},
        title = "{NICER Observation of the Temporal and Spectral Evolution of Swift J1818.0-1607: A Missing Link between Magnetars and Rotation-powered Pulsars}",
      journal = {\apj},
         year = 2020,
        month = oct,
       volume = {902},
       number = {1},
          eid = {1},
        pages = {1},
          doi = {10.3847/1538-4357/abb3c9},
archivePrefix = {arXiv},
       eprint = {2009.00231},
 primaryClass = {astro-ph.HE},
       adsurl = {https://ui.adsabs.harvard.edu/abs/2020ApJ...902....1H}
}

@ARTICLE{Beniamini2019,
       author = {{Beniamini}, Paz and {Hotokezaka}, Kenta and {van der Horst}, Alexander and {Kouveliotou}, Chryssa},
        title = "{Formation rates and evolution histories of magnetars}",
      journal = {\mnras},
         year = 2019,
        month = jul,
       volume = {487},
       number = {1},
        pages = {1426-1438},
          doi = {10.1093/mnras/stz1391},
archivePrefix = {arXiv},
       eprint = {1903.06718},
 primaryClass = {astro-ph.HE},
       adsurl = {https://ui.adsabs.harvard.edu/abs/2019MNRAS.487.1426B}
}

@ARTICLE{Lazarus2012,
       author = {{Lazarus}, P. and {Kaspi}, V.~M. and {Champion}, D.~J. and {Hessels}, J.~W.~T. and {Dib}, R.},
        title = "{Constraining Radio Emission from Magnetars}",
      journal = {\apj},
         year = 2012,
        month = jan,
       volume = {744},
       number = {2},
          eid = {97},
        pages = {97},
          doi = {10.1088/0004-637X/744/2/97},
archivePrefix = {arXiv},
       eprint = {1109.5116},
 primaryClass = {astro-ph.HE},
       adsurl = {https://ui.adsabs.harvard.edu/abs/2012ApJ...744...97L}
}

@ARTICLE{Majid2011,
       author = {{Majid}, Walid A. and {Naudet}, Charles J. and {Lowe}, Stephen T. and {Kuiper}, Thomas B.~H.},
        title = "{Statistical Studies of Giant Pulse Emission from the Crab Pulsar}",
      journal = {\apj},
         year = 2011,
        month = nov,
       volume = {741},
       number = {1},
          eid = {53},
        pages = {53},
          doi = {10.1088/0004-637X/741/1/53},
archivePrefix = {arXiv},
       eprint = {1108.2307},
 primaryClass = {astro-ph.GA},
       adsurl = {https://ui.adsabs.harvard.edu/abs/2011ApJ...741...53M}
}

@ARTICLE{Bera2019,
       author = {{Bera}, Apurba and {Chengalur}, Jayaram N.},
        title = "{Super-giant pulses from the Crab pulsar: energy distribution and occurrence rate}",
      journal = {\mnras},
         year = 2019,
        month = nov,
       volume = {490},
       number = {1},
        pages = {L12-L16},
          doi = {10.1093/mnrasl/slz140},
archivePrefix = {arXiv},
       eprint = {1909.13812},
 primaryClass = {astro-ph.HE},
       adsurl = {https://ui.adsabs.harvard.edu/abs/2019MNRAS.490L..12B}
}

@ARTICLE{Karuppusamy2010,
       author = {{Karuppusamy}, R. and {Stappers}, B.~W. and {van Straten}, W.},
        title = "{Giant pulses from the Crab pulsar. A wide-band study}",
      journal = {\aap},
         year = 2010,
        month = jun,
       volume = {515},
          eid = {A36},
        pages = {A36},
          doi = {10.1051/0004-6361/200913729},
archivePrefix = {arXiv},
       eprint = {1004.2803},
 primaryClass = {astro-ph.GA},
       adsurl = {https://ui.adsabs.harvard.edu/abs/2010A&A...515A..36K}
}

@ARTICLE{Lyne1993,
       author = {{Lyne}, A.~G. and {Pritchard}, R.~S. and {Graham Smith}, F.},
        title = "{23 years of Crab pulsar rotational history.}",
      journal = {\mnras},
         year = 1993,
        month = dec,
       volume = {265},
        pages = {1003-1012},
          doi = {10.1093/mnras/265.4.1003},
       adsurl = {https://ui.adsabs.harvard.edu/abs/1993MNRAS.265.1003L}
}

@ARTICLE{Lewandowska2022,
       author = {{Lewandowska}, N. and {Demorest}, P.~B. and {McLaughlin}, M.~A. and {Kilian}, P. and {Hankins}, T.~H.},
        title = "{Single Pulse Dispersion Measure of the Crab Pulsar}",
      journal = {\apj},
         year = 2022,
        month = aug,
       volume = {935},
       number = {2},
          eid = {84},
        pages = {84},
          doi = {10.3847/1538-4357/ac8055},
archivePrefix = {arXiv},
       eprint = {2207.04267},
 primaryClass = {astro-ph.SR},
       adsurl = {https://ui.adsabs.harvard.edu/abs/2022ApJ...935...84L}
}

@ARTICLE{McKee2018,
       author = {{McKee}, J.~W. and {Lyne}, A.~G. and {Stappers}, B.~W. and {Bassa}, C.~G. and {Jordan}, C.~A.},
        title = "{Temporal variations in scattering and dispersion measure in the Crab Pulsar and their effect on timing precision}",
      journal = {\mnras},
         year = 2018,
        month = sep,
       volume = {479},
       number = {3},
        pages = {4216-4224},
          doi = {10.1093/mnras/sty1727},
archivePrefix = {arXiv},
       eprint = {1806.10486},
 primaryClass = {astro-ph.HE},
       adsurl = {https://ui.adsabs.harvard.edu/abs/2018MNRAS.479.4216M}
}

@ARTICLE{Margalit2020,
       author = {{Margalit}, Ben and {Beniamini}, Paz and {Sridhar}, Navin and {Metzger}, Brian D.},
        title = "{Implications of a Fast Radio Burst from a Galactic Magnetar}",
      journal = {\apjl},
         year = 2020,
        month = aug,
       volume = {899},
       number = {2},
          eid = {L27},
        pages = {L27},
          doi = {10.3847/2041-8213/abac57},
archivePrefix = {arXiv},
       eprint = {2005.05283},
 primaryClass = {astro-ph.HE},
       adsurl = {https://ui.adsabs.harvard.edu/abs/2020ApJ...899L..27M}
}

@ARTICLE{Beloborodov2020,
       author = {{Beloborodov}, Andrei M.},
        title = "{Blast Waves from Magnetar Flares and Fast Radio Bursts}",
      journal = {\apj},
         year = 2020,
        month = jun,
       volume = {896},
       number = {2},
          eid = {142},
        pages = {142},
          doi = {10.3847/1538-4357/ab83eb},
archivePrefix = {arXiv},
       eprint = {1908.07743},
 primaryClass = {astro-ph.HE},
       adsurl = {https://ui.adsabs.harvard.edu/abs/2020ApJ...896..142B}
}

@ARTICLE{Metzger2019,
       author = {{Metzger}, Brian D. and {Margalit}, Ben and {Sironi}, Lorenzo},
        title = "{Fast radio bursts as synchrotron maser emission from decelerating relativistic blast waves}",
      journal = {\mnras},
         year = 2019,
        month = may,
       volume = {485},
       number = {3},
        pages = {4091-4106},
          doi = {10.1093/mnras/stz700},
archivePrefix = {arXiv},
       eprint = {1902.01866},
 primaryClass = {astro-ph.HE},
       adsurl = {https://ui.adsabs.harvard.edu/abs/2019MNRAS.485.4091M}
}

@ARTICLE{Kumar2017,
       author = {{Kumar}, Pawan and {Lu}, Wenbin and {Bhattacharya}, Mukul},
        title = "{Fast radio burst source properties and curvature radiation model}",
      journal = {\mnras},
         year = 2017,
        month = jul,
       volume = {468},
       number = {3},
        pages = {2726-2739},
          doi = {10.1093/mnras/stx665},
archivePrefix = {arXiv},
       eprint = {1703.06139},
 primaryClass = {astro-ph.HE},
       adsurl = {https://ui.adsabs.harvard.edu/abs/2017MNRAS.468.2726K}
}

@ARTICLE{Kirsten2024,
       author = {{Kirsten}, F. and {Ould-Boukattine}, O.~S. and {Herrmann}, W. and {Gawro{\'n}ski}, M.~P. and {Hessels}, J.~W.~T. and {Lu}, W. and {Snelders}, M.~P. and {Chawla}, P. and {Yang}, J. and {Blaauw}, R. and {Nimmo}, K. and {Puchalska}, W. and {Wolak}, P. and {van Ruiten}, R.},
        title = "{A link between repeating and non-repeating fast radio bursts through their energy distributions}",
      journal = {Nature Astronomy},
         year = 2024,
        month = mar,
       volume = {8},
        pages = {337-346},
          doi = {10.1038/s41550-023-02153-z},
archivePrefix = {arXiv},
       eprint = {2306.15505},
 primaryClass = {astro-ph.HE},
       adsurl = {https://ui.adsabs.harvard.edu/abs/2024NatAs...8..337K}
}

@ARTICLE{CHIME_SGR2022b,
       author = {{Pearlman}, Aaron B. and {Chime/Frb Collaboration}},
        title = "{CHIME/FRB Detection of Another Bright Radio Burst from SGR 1935+2154}",
      journal = {The Astronomer's Telegram},
         year = 2022,
        month = dec,
       volume = {15792},
        pages = {1},
       adsurl = {https://ui.adsabs.harvard.edu/abs/2022ATel15792....1P}
}

@ARTICLE{Yunnan_SGR2022,
       author = {{Huang}, Y.~X. and {Xu}, H. and {Xu}, Y.~H. and {Wang}, B.~J. and {Li}, Z.~X. and {Zhang}, C.~F. and {Xu}, J.~W. and {Jiang}, J.~C. and {Men}, Y.~P. and {Gao}, G.~N. and {Xu}, Z.~H. and {Hao}, L.~F. and {Lee}, K.~J. and {Wang}, M. and {Zhang}, B. and {Zhu}, W.~W.},
        title = "{An intermediately bright radio burst detected from SGR 1935+2154 in S-band by Yunnan 40m radio telescope}",
      journal = {The Astronomer's Telegram},
         year = 2022,
        month = oct,
       volume = {15707},
        pages = {1},
       adsurl = {https://ui.adsabs.harvard.edu/abs/2022ATel15707....1H}
}

@ARTICLE{FASTR_SGR2022,
       author = {{Zhu}, Weiwei and {Xu}, Heng and {Zhou}, Dejiang and {Lin}, Lin and {Wang}, Bojun and {Wang}, Pei and {Zhang}, Chunfeng and {Niu}, Jiarui and {Chen}, Yutong and {Li}, Chengkui and {Meng}, Lingqi and {Lee}, Kejia and {Zhang}, Bing and {Feng}, Yi and {Ge}, Mingyu and {G{\"o}{\u{g}}{\"u}{\textcommabelow s}}, Ersin and {Guan}, Xing and {Han}, Jinlin and {Jiang}, Jinchen and {Jiang}, Peng and {Kouveliotou}, Chryssa and {Li}, Di and {Miao}, Chenchen and {Miao}, Xueli and {Men}, Yunpeng and {Niu}, Chenghui and {Wang}, Weiyang and {Wang}, Zhengli and {Xu}, Jiangwei and {Xu}, Renxin and {Xue}, Mengyao and {Yang}, Yuanpei and {Yu}, Wenfei and {Yuan}, Mao and {Yue}, Youling and {Zhang}, Shuangnan and {Zhang}, Yongkun},
        title = "{A radio pulsar phase from SGR J1935+2154 provides clues to the magnetar FRB mechanism}",
      journal = {Science Advances},
         year = 2023,
        month = jul,
       volume = {9},
       number = {30},
          eid = {eadf6198},
        pages = {eadf6198},
          doi = {10.1126/sciadv.adf6198},
archivePrefix = {arXiv},
       eprint = {2307.16124},
 primaryClass = {astro-ph.HE},
       adsurl = {https://ui.adsabs.harvard.edu/abs/2023SciA....9F6198Z}
}

@ARTICLE{GBT_SGR2022,
       author = {{Maan}, Yogesh and {Leeuwen}, Joeri van and {Straal}, Samayra and {Pastor-Marazuela}, Ines},
        title = "{GBT detection of bright 5 GHz radio bursts from SGR 1935+2154, coincident with X-ray and 600 MHz bursts}",
      journal = {The Astronomer's Telegram},
         year = 2022,
        month = oct,
       volume = {15697},
        pages = {1},
       adsurl = {https://ui.adsabs.harvard.edu/abs/2022ATel15697....1M}
}

@ARTICLE{CHIME_SGR2022,
       author = {{Dong}, Fengqiu Adam and {Chime/Frb Collaboration}},
        title = "{CHIME/FRB Detection of a Bright Radio Burst from SGR 1935+2154}",
      journal = {The Astronomer's Telegram},
         year = 2022,
        month = oct,
       volume = {15681},
        pages = {1},
       adsurl = {https://ui.adsabs.harvard.edu/abs/2022ATel15681....1D}
}

@ARTICLE{FAST_SGR2020,
       author = {{Zhang}, C.~F. and {Jiang}, J.~C. and {Men}, Y.~P. and {Wang}, B.~J. and {Xu}, H. and {Xu}, J.~W. and {Niu}, C.~H. and {Zhou}, D.~J. and {Guan}, X. and {Han}, J.~L. and {Jiang}, P. and {Lee}, K.~J. and {Li}, D. and {Lin}, L. and {Niu}, J.~R. and {Wang}, P. and {Wang}, Z.~L. and {Xu}, R.~X. and {Yu}, W. and {Zhang}, B. and {Zhu}, W.~W.},
        title = "{A highly polarised radio burst detected from SGR 1935+2154 by FAST}",
      journal = {The Astronomer's Telegram},
         year = 2020,
        month = may,
       volume = {13699},
        pages = {1},
       adsurl = {https://ui.adsabs.harvard.edu/abs/2020ATel13699....1Z}
}

@ARTICLE{Li2021,
       author = {{Li}, C.~K. and {Lin}, L. and {Xiong}, S.~L. and {Ge}, M.~Y. and {Li}, X.~B. and {Li}, T.~P. and {Lu}, F.~J. and {Zhang}, S.~N. and {Tuo}, Y.~L. and {Nang}, Y. and {Zhang}, B. and {Xiao}, S. and {Chen}, Y. and {Song}, L.~M. and {Xu}, Y.~P. and {Liu}, C.~Z. and {Jia}, S.~M. and {Cao}, X.~L. and {Qu}, J.~L. and {Zhang}, S. and {Gu}, Y.~D. and {Liao}, J.~Y. and {Zhao}, X.~F. and {Tan}, Y. and {Nie}, J.~Y. and {Zhao}, H.~S. and {Zheng}, S.~J. and {Zheng}, Y.~G. and {Luo}, Q. and {Cai}, C. and {Li}, B. and {Xue}, W.~C. and {Bu}, Q.~C. and {Chang}, Z. and {Chen}, G. and {Chen}, L. and {Chen}, T.~X. and {Chen}, Y.~B. and {Chen}, Y.~P. and {Cui}, W. and {Cui}, W.~W. and {Deng}, J.~K. and {Dong}, Y.~W. and {Du}, Y.~Y. and {Fu}, M.~X. and {Gao}, G.~H. and {Gao}, H. and {Gao}, M. and {Gu}, Y.~D. and {Guan}, J. and {Guo}, C.~C. and {Han}, D.~W. and {Huang}, Y. and {Huo}, J. and {Jiang}, L.~H. and {Jiang}, W.~C. and {Jin}, J. and {Jin}, Y.~J. and {Kong}, L.~D. and {Li}, G. and {Li}, M.~S. and {Li}, W. and {Li}, X. and {Li}, X.~F. and {Li}, Y.~G. and {Li}, Z.~W. and {Liang}, X.~H. and {Liu}, B.~S. and {Liu}, G.~Q. and {Liu}, H.~W. and {Liu}, X.~J. and {Liu}, Y.~N. and {Lu}, B. and {Lu}, X.~F. and {Luo}, T. and {Ma}, X. and {Meng}, B. and {Ou}, G. and {Sai}, N. and {Shang}, R.~C. and {Song}, X.~Y. and {Sun}, L. and {Tao}, L. and {Wang}, C. and {Wang}, G.~F. and {Wang}, J. and {Wang}, W.~S. and {Wang}, Y.~S. and {Wen}, X.~Y. and {Wu}, B.~B. and {Wu}, B.~Y. and {Wu}, M. and {Xiao}, G.~C. and {Xu}, H. and {Yang}, J.~W. and {Yang}, S. and {Yang}, Y.~J. and {Yang}, Yi-Jung and {Yi}, Q.~B. and {Yin}, Q.~Q. and {You}, Y. and {Zhang}, A.~M. and {Zhang}, C.~M. and {Zhang}, F. and {Zhang}, H.~M. and {Zhang}, J. and {Zhang}, T. and {Zhang}, W. and {Zhang}, W.~C. and {Zhang}, W.~Z. and {Zhang}, Y. and {Zhang}, Yue and {Zhang}, Y.~F. and {Zhang}, Y.~J. and {Zhang}, Z. and {Zhang}, Zhi and {Zhang}, Z.~L. and {Zhou}, D.~K. and {Zhou}, J.~F. and {Zhu}, Y. and {Zhu}, Y.~X. and {Zhuang}, R.~L.},
        title = "{HXMT identification of a non-thermal X-ray burst from SGR J1935+2154 and with FRB 200428}",
      journal = {Nature Astronomy},
         year = 2021,
        month = apr,
       volume = {5},
        pages = {378-384},
          doi = {10.1038/s41550-021-01302-6},
archivePrefix = {arXiv},
       eprint = {2005.11071},
 primaryClass = {astro-ph.HE},
       adsurl = {https://ui.adsabs.harvard.edu/abs/2021NatAs...5..378L}
}

@ARTICLE{Ridnaia2020,
       author = {{Ridnaia}, A. and {Svinkin}, D. and {Frederiks}, D. and {Bykov}, A. and {Popov}, S. and {Aptekar}, R. and {Golenetskii}, S. and {Lysenko}, A. and {Tsvetkova}, A. and {Ulanov}, M. and {Cline}, T.~L.},
        title = "{A peculiar hard X-ray counterpart of a Galactic fast radio burst}",
      journal = {Nature Astronomy},
         year = 2021,
        month = apr,
       volume = {5},
        pages = {372-377},
          doi = {10.1038/s41550-020-01265-0},
archivePrefix = {arXiv},
       eprint = {2005.11178},
 primaryClass = {astro-ph.HE},
       adsurl = {https://ui.adsabs.harvard.edu/abs/2021NatAs...5..372R}
}

@ARTICLE{Mereghetti2020,
       author = {{Mereghetti}, S. and {Savchenko}, V. and {Ferrigno}, C. and {G{\"o}tz}, D. and {Rigoselli}, M. and {Tiengo}, A. and {Bazzano}, A. and {Bozzo}, E. and {Coleiro}, A. and {Courvoisier}, T.~J.-L. and {Doyle}, M. and {Goldwurm}, A. and {Hanlon}, L. and {Jourdain}, E. and {von Kienlin}, A. and {Lutovinov}, A. and {Martin-Carrillo}, A. and {Molkov}, S. and {Natalucci}, L. and {Onori}, F. and {Panessa}, F. and {Rodi}, J. and {Rodriguez}, J. and {S{\'a}nchez-Fern{\'a}ndez}, C. and {Sunyaev}, R. and {Ubertini}, P.},
        title = "{INTEGRAL Discovery of a Burst with Associated Radio Emission from the Magnetar SGR 1935+2154}",
      journal = {\apjl},
         year = 2020,
        month = aug,
       volume = {898},
       number = {2},
          eid = {L29},
        pages = {L29},
          doi = {10.3847/2041-8213/aba2cf},
archivePrefix = {arXiv},
       eprint = {2005.06335},
 primaryClass = {astro-ph.HE},
       adsurl = {https://ui.adsabs.harvard.edu/abs/2020ApJ...898L..29M}
}

@ARTICLE{CHIME_SGR2020,
       author = {{CHIME/FRB Collaboration} and {Andersen}, B.~C. and {Bandura}, K.~M. and {Bhardwaj}, M. and {Bij}, A. and {Boyce}, M.~M. and {Boyle}, P.~J. and {Brar}, C. and {Cassanelli}, T. and {Chawla}, P. and {Chen}, T. and {Cliche}, J.-F. and {Cook}, A. and {Cubranic}, D. and {Curtin}, A.~P. and {Denman}, N.~T. and {Dobbs}, M. and {Dong}, F.~Q. and {Fandino}, M. and {Fonseca}, E. and {Gaensler}, B.~M. and {Giri}, U. and {Good}, D.~C. and {Halpern}, M. and {Hill}, A.~S. and {Hinshaw}, G.~F. and {H{\"o}fer}, C. and {Josephy}, A. and {Kania}, J.~W. and {Kaspi}, V.~M. and {Landecker}, T.~L. and {Leung}, C. and {Li}, D.~Z. and {Lin}, H.-H. and {Masui}, K.~W. and {McKinven}, R. and {Mena-Parra}, J. and {Merryfield}, M. and {Meyers}, B.~W. and {Michilli}, D. and {Milutinovic}, N. and {Mirhosseini}, A. and {M{\"u}nchmeyer}, M. and {Naidu}, A. and {Newburgh}, L.~B. and {Ng}, C. and {Patel}, C. and {Pen}, U.-L. and {Pinsonneault-Marotte}, T. and {Pleunis}, Z. and {Quine}, B.~M. and {Rafiei-Ravandi}, M. and {Rahman}, M. and {Ransom}, S.~M. and {Renard}, A. and {Sanghavi}, P. and {Scholz}, P. and {Shaw}, J.~R. and {Shin}, K. and {Siegel}, S.~R. and {Singh}, S. and {Smegal}, R.~J. and {Smith}, K.~M. and {Stairs}, I.~H. and {Tan}, C.~M. and {Tendulkar}, S.~P. and {Tretyakov}, I. and {Vanderlinde}, K. and {Wang}, H. and {Wulf}, D. and {Zwaniga}, A.~V.},
        title = "{A bright millisecond-duration radio burst from a Galactic magnetar}",
      journal = {\nat},
         year = 2020,
        month = nov,
       volume = {587},
       number = {7832},
        pages = {54-58},
          doi = {10.1038/s41586-020-2863-y},
archivePrefix = {arXiv},
       eprint = {2005.10324},
 primaryClass = {astro-ph.HE},
       adsurl = {https://ui.adsabs.harvard.edu/abs/2020Natur.587...54C}
}

@ARTICLE{Oostrum2020,
       author = {{Oostrum}, L.~C. and {Maan}, Y. and {van Leeuwen}, J. and {Connor}, L. and {Petroff}, E. and {Attema}, J.~J. and {Bast}, J.~E. and {Gardenier}, D.~W. and {Hargreaves}, J.~E. and {Kooistra}, E. and {van der Schuur}, D. and {Sclocco}, A. and {Smits}, R. and {Straal}, S.~M. and {ter Veen}, S. and {Vohl}, D. and {Adams}, E.~A.~K. and {Adebahr}, B. and {de Blok}, W.~J.~G. and {van den Brink}, R.~H. and {van Cappellen}, W.~A. and {Coolen}, A.~H.~W.~M. and {Damstra}, S. and {van Diepen}, G.~N.~J. and {Frank}, B.~S. and {Hess}, K.~M. and {van der Hulst}, J.~M. and {Hut}, B. and {Ivashina}, M.~V. and {Loose}, G.~M. and {Lucero}, D.~M. and {Mika}, {\'A}. and {Morganti}, R.~H. and {Moss}, V.~A. and {Mulder}, H. and {Norden}, M.~J. and {Oosterloo}, T.~A. and {Orr{\'u}}, E. and {de Reijer}, J.~P.~R. and {Ruiter}, M. and {Vermaas}, N.~J. and {Wijnholds}, S.~J. and {Ziemke}, J.},
        title = "{Repeating fast radio bursts with WSRT/Apertif}",
      journal = {\aap},
         year = 2020,
        month = mar,
       volume = {635},
          eid = {A61},
        pages = {A61},
          doi = {10.1051/0004-6361/201937422},
archivePrefix = {arXiv},
       eprint = {1912.12217},
 primaryClass = {astro-ph.HE},
       adsurl = {https://ui.adsabs.harvard.edu/abs/2020A&A...635A..61O}
}

@INPROCEEDINGS{Apertif2017,
       author = {{Maan}, Yogesh and {van Leeuwen}, Joeri},
        title = "{Real-time searches for fast transients with Apertif and LOFAR}",
    booktitle = {2017 XXXIInd General Assembly and Scientific Symposium of the International Union of Radio Science (URSI GASS},
         year = 2017,
        month = sep,
          eid = {2},
        pages = {2},
          doi = {10.23919/URSIGASS.2017.8105320},
archivePrefix = {arXiv},
       eprint = {1709.06104},
 primaryClass = {astro-ph.IM},
       adsurl = {https://ui.adsabs.harvard.edu/abs/2017ursi.confE...2M}
}

@ARTICLE{Bhandari2018,
       author = {{Bhandari}, S. and {Keane}, E.~F. and {Barr}, E.~D. and {Jameson}, A. and {Petroff}, E. and {Johnston}, S. and {Bailes}, M. and {Bhat}, N.~D.~R. and {Burgay}, M. and {Burke-Spolaor}, S. and {Caleb}, M. and {Eatough}, R.~P. and {Flynn}, C. and {Green}, J.~A. and {Jankowski}, F. and {Kramer}, M. and {Krishnan}, V. Venkatraman and {Morello}, V. and {Possenti}, A. and {Stappers}, B. and {Tiburzi}, C. and {van Straten}, W. and {Andreoni}, I. and {Butterley}, T. and {Chandra}, P. and {Cooke}, J. and {Corongiu}, A. and {Coward}, D.~M. and {Dhillon}, V.~S. and {Dodson}, R. and {Hardy}, L.~K. and {Howell}, E.~J. and {Jaroenjittichai}, P. and {Klotz}, A. and {Littlefair}, S.~P. and {Marsh}, T.~R. and {Mickaliger}, M. and {Muxlow}, T. and {Perrodin}, D. and {Pritchard}, T. and {Sawangwit}, U. and {Terai}, T. and {Tominaga}, N. and {Torne}, P. and {Totani}, T. and {Trois}, A. and {Turpin}, D. and {Niino}, Y. and {Wilson}, R.~W. and {Albert}, A. and {Andr{\'e}}, M. and {Anghinolfi}, M. and {Anton}, G. and {Ardid}, M. and {Aubert}, J.-J. and {Avgitas}, T. and {Baret}, B. and {Barrios-Mart{\'\i}}, J. and {Basa}, S. and {Belhorma}, B. and {Bertin}, V. and {Biagi}, S. and {Bormuth}, R. and {Bourret}, S. and {Bouwhuis}, M.~C. and {Br{\^a}nza{\c{s}}}, H. and {Bruijn}, R. and {Brunner}, J. and {Busto}, J. and {Capone}, A. and {Caramete}, L. and {Carr}, J. and {Celli}, S. and {Moursli}, R. Cherkaoui El and {Chiarusi}, T. and {Circella}, M. and {Coelho}, J.~A.~B. and {Coleiro}, A. and {Coniglione}, R. and {Costantini}, H. and {Coyle}, P. and {Creusot}, A. and {D{\'\i}az}, A.~F. and {Deschamps}, A. and {De Bonis}, G. and {Distefano}, C. and {Palma}, I. Di and {Domi}, A. and {Donzaud}, C. and {Dornic}, D. and {Drouhin}, D. and {Eberl}, T. and {Bojaddaini}, I. El and {Khayati}, N. El and {Els{\"a}sser}, D. and {Enzenh{\"o}fer}, A. and {Ettahiri}, A. and {Fassi}, F. and {Felis}, I. and {Fusco}, L.~A. and {Gay}, P. and {Giordano}, V. and {Glotin}, H. and {Gregoire}, T. and {Gracia-Ruiz}, R. and {Graf}, K. and {Hallmann}, S. and {van Haren}, H. and {Heijboer}, A.~J. and {Hello}, Y. and {Hern{\'a}ndez-Rey}, J.~J. and {H{\"o}{\ss}l}, J. and {Hofest{\"a}dt}, J. and {Hugon}, C. and {Illuminati}, G. and {James}, C.~W. and {de Jong}, M. and {Jongen}, M. and {Kadler}, M. and {Kalekin}, O. and {Katz}, U. and {Kie{\ss}ling}, D. and {Kouchner}, A. and {Kreter}, M. and {Kreykenbohm}, I. and {Kulikovskiy}, V. and {Lachaud}, C. and {Lahmann}, R. and {Lef{\`e}vre}, D. and {Leonora}, E. and {Loucatos}, S. and {Marcelin}, M. and {Margiotta}, A. and {Marinelli}, A. and {Mart{\'\i}nez-Mora}, J.~A. and {Mele}, R. and {Melis}, K. and {Michael}, T. and {Migliozzi}, P. and {Moussa}, A. and {Navas}, S. and {Nezri}, E. and {Organokov}, M. and {P{\v{a}}v{\v{a}}la{\c{s}}}, G.~E. and {Pellegrino}, C. and {Perrina}, C. and {Piattelli}, P. and {Popa}, V. and {Pradier}, T. and {Quinn}, L. and {Racca}, C. and {Riccobene}, G. and {S{\'a}nchez-Losa}, A. and {Salda{\~n}a}, M. and {Salvadori}, I. and {Samtleben}, D.~F.~E. and {Sanguineti}, M. and {Sapienza}, P. and {Sch{\"u}ssler}, F. and {Sieger}, C. and {Spurio}, M. and {Stolarczyk}, Th and {Taiuti}, M. and {Tayalati}, Y. and {Trovato}, A. and {Turpin}, D. and {T{\"o}nnis}, C. and {Vallage}, B. and {Van Elewyck}, V. and {Versari}, F. and {Vivolo}, D. and {Vizzocca}, A. and {Wilms}, J. and {Zornoza}, J.~D. and {Z{\'u}{\~n}iga}, J.},
        title = "{The SUrvey for Pulsars and Extragalactic Radio Bursts - II. New FRB discoveries and their follow-up}",
      journal = {\mnras},
         year = 2018,
        month = apr,
       volume = {475},
       number = {2},
        pages = {1427-1446},
          doi = {10.1093/mnras/stx3074},
archivePrefix = {arXiv},
       eprint = {1711.08110},
 primaryClass = {astro-ph.HE},
       adsurl = {https://ui.adsabs.harvard.edu/abs/2018MNRAS.475.1427B}
}

@ARTICLE{Champion2016,
       author = {{Champion}, D.~J. and {Petroff}, E. and {Kramer}, M. and {Keith}, M.~J. and {Bailes}, M. and {Barr}, E.~D. and {Bates}, S.~D. and {Bhat}, N.~D.~R. and {Burgay}, M. and {Burke-Spolaor}, S. and {Flynn}, C.~M.~L. and {Jameson}, A. and {Johnston}, S. and {Ng}, C. and {Levin}, L. and {Possenti}, A. and {Stappers}, B.~W. and {van Straten}, W. and {Thornton}, D. and {Tiburzi}, C. and {Lyne}, A.~G.},
        title = "{Five new fast radio bursts from the HTRU high-latitude survey at Parkes: first evidence for two-component bursts}",
      journal = {\mnras},
         year = 2016,
        month = jul,
       volume = {460},
       number = {1},
        pages = {L30-L34},
          doi = {10.1093/mnrasl/slw069},
archivePrefix = {arXiv},
       eprint = {1511.07746},
 primaryClass = {astro-ph.HE},
       adsurl = {https://ui.adsabs.harvard.edu/abs/2016MNRAS.460L..30C}
}

@ARTICLE{ATA2009,
       author = {{Welch}, J. and {Backer}, D. and {Blitz}, L. and {Bock}, D.~C.-J. and {Bower}, G.~C. and {Cheng}, C. and {Croft}, S. and {Dexter}, M. and {Engargiola}, G. and {Fields}, E. and {Forster}, J. and {Gutierrez-Kraybill}, C. and {Heiles}, C. and {Helfer}, T. and {Jorgensen}, S. and {Keating}, G. and {Lugten}, J. and {MacMahon}, D. and {Milgrome}, O. and {Thornton}, D. and {Urry}, L. and {van Leeuwen}, J. and {Werthimer}, D. and {Williams}, P.~H. and {Wright}, M. and {Tarter}, J. and {Ackermann}, R. and {Atkinson}, S. and {Backus}, P. and {Barott}, W. and {Bradford}, T. and {Davis}, M. and {Deboer}, D. and {Dreher}, J. and {Harp}, G. and {Jordan}, J. and {Kilsdonk}, T. and {Pierson}, T. and {Randall}, K. and {Ross}, J. and {Shostak}, S. and {Fleming}, M. and {Cork}, C. and {Vitouchkine}, A. and {Wadefalk}, N. and {Weinreb}, S.},
        title = "{The Allen Telescope Array: The First Widefield, Panchromatic, Snapshot Radio Camera for Radio Astronomy and SETI}",
      journal = {IEEE Proceedings},
         year = 2009,
        month = aug,
       volume = {97},
       number = {8},
        pages = {1438-1447},
          doi = {10.1109/JPROC.2009.2017103},
archivePrefix = {arXiv},
       eprint = {0904.0762},
 primaryClass = {astro-ph.IM},
       adsurl = {https://ui.adsabs.harvard.edu/abs/2009IEEEP..97.1438W}
}

@ARTICLE{Sheikh2024,
       author = {{Sheikh}, Sofia Z. and {Farah}, Wael and {Pollak}, Alexander W. and {Siemion}, Andrew P.~V. and {Chamma}, Mohammed A. and {Cruz}, Luigi F. and {Davis}, Roy H. and {DeBoer}, David R. and {Gajjar}, Vishal and {Karn}, Phil and {Kittling}, Jamar and {Lu}, Wenbin and {Masters}, Mark and {Premnath}, Pranav and {Schoultz}, Sarah and {Shumaker}, Carol and {Singh}, Gurmehar and {Snodgrass}, Michael},
        title = "{Characterization of the repeating FRB 20220912A with the Allen Telescope Array}",
      journal = {\mnras},
         year = 2024,
        month = feb,
       volume = {527},
       number = {4},
        pages = {10425-10439},
          doi = {10.1093/mnras/stad3630},
archivePrefix = {arXiv},
       eprint = {2312.07756},
 primaryClass = {astro-ph.HE},
       adsurl = {https://ui.adsabs.harvard.edu/abs/2024MNRAS.52710425S}
}

@ARTICLE{STARE2_2020,
       author = {{Bochenek}, Christopher D. and {McKenna}, Daniel L. and {Belov}, Konstantin V. and {Kocz}, Jonathon and {Kulkarni}, S.~R. and {Lamb}, James and {Ravi}, Vikram and {Woody}, David},
        title = "{STARE2: Detecting Fast Radio Bursts in the Milky Way}",
      journal = {\pasp},
         year = 2020,
        month = mar,
       volume = {132},
       number = {1009},
          eid = {034202},
        pages = {034202},
          doi = {10.1088/1538-3873/ab63b3},
archivePrefix = {arXiv},
       eprint = {2001.05077},
 primaryClass = {astro-ph.HE},
       adsurl = {https://ui.adsabs.harvard.edu/abs/2020PASP..132c4202B}
}

@ARTICLE{Caleb2017,
       author = {{Caleb}, M. and {Flynn}, C. and {Bailes}, M. and {Barr}, E.~D. and {Bateman}, T. and {Bhandari}, S. and {Campbell-Wilson}, D. and {Farah}, W. and {Green}, A.~J. and {Hunstead}, R.~W. and {Jameson}, A. and {Jankowski}, F. and {Keane}, E.~F. and {Parthasarathy}, A. and {Ravi}, V. and {Rosado}, P.~A. and {van Straten}, W. and {Venkatraman Krishnan}, V.},
        title = "{The first interferometric detections of fast radio bursts}",
      journal = {\mnras},
         year = 2017,
        month = jul,
       volume = {468},
       number = {3},
        pages = {3746-3756},
          doi = {10.1093/mnras/stx638},
archivePrefix = {arXiv},
       eprint = {1703.10173},
 primaryClass = {astro-ph.HE},
       adsurl = {https://ui.adsabs.harvard.edu/abs/2017MNRAS.468.3746C}
}

@ARTICLE{Bailes2017,
       author = {{Bailes}, M. and {Jameson}, A. and {Flynn}, C. and {Bateman}, T. and {Barr}, E.~D. and {Bhandari}, S. and {Bunton}, J.~D. and {Caleb}, M. and {Campbell-Wilson}, D. and {Farah}, W. and {Gaensler}, B. and {Green}, A.~J. and {Hunstead}, R.~W. and {Jankowski}, F. and {Keane}, E.~F. and {Krishnan}, V. Venkatraman and {Murphy}, Tara and {O'Neill}, M. and {Os{\l}owski}, S. and {Parthasarathy}, A. and {Ravi}, V. and {Rosado}, P. and {Temby}, D.},
        title = "{The UTMOST: A Hybrid Digital Signal Processor Transforms the Molonglo Observatory Synthesis Telescope}",
      journal = {\pasa},
         year = 2017,
        month = oct,
       volume = {34},
          eid = {e045},
        pages = {e045},
          doi = {10.1017/pasa.2017.39},
archivePrefix = {arXiv},
       eprint = {1708.09619},
 primaryClass = {astro-ph.IM},
       adsurl = {https://ui.adsabs.harvard.edu/abs/2017PASA...34...45B}
}

@ARTICLE{Shannon2018,
       author = {{Shannon}, R.~M. and {Macquart}, J.-P. and {Bannister}, K.~W. and {Ekers}, R.~D. and {James}, C.~W. and {Os{\l}owski}, S. and {Qiu}, H. and {Sammons}, M. and {Hotan}, A.~W. and {Voronkov}, M.~A. and {Beresford}, R.~J. and {Brothers}, M. and {Brown}, A.~J. and {Bunton}, J.~D. and {Chippendale}, A.~P. and {Haskins}, C. and {Leach}, M. and {Marquarding}, M. and {McConnell}, D. and {Pilawa}, M.~A. and {Sadler}, E.~M. and {Troup}, E.~R. and {Tuthill}, J. and {Whiting}, M.~T. and {Allison}, J.~R. and {Anderson}, C.~S. and {Bell}, M.~E. and {Collier}, J.~D. and {G{\"u}rkan}, G. and {Heald}, G. and {Riseley}, C.~J.},
        title = "{The dispersion-brightness relation for fast radio bursts from a wide-field survey}",
      journal = {\nat},
         year = 2018,
        month = oct,
       volume = {562},
       number = {7727},
        pages = {386-390},
          doi = {10.1038/s41586-018-0588-y},
       adsurl = {https://ui.adsabs.harvard.edu/abs/2018Natur.562..386S}
}

@ARTICLE{Bannister2017,
       author = {{Bannister}, K.~W. and {Shannon}, R.~M. and {Macquart}, J.-P. and {Flynn}, C. and {Edwards}, P.~G. and {O'Neill}, M. and {Os{\l}owski}, S. and {Bailes}, M. and {Zackay}, B. and {Clarke}, N. and {D'Addario}, L.~R. and {Dodson}, R. and {Hall}, P.~J. and {Jameson}, A. and {Jones}, D. and {Navarro}, R. and {Trinh}, J.~T. and {Allison}, J. and {Anderson}, C.~S. and {Bell}, M. and {Chippendale}, A.~P. and {Collier}, J.~D. and {Heald}, G. and {Heywood}, I. and {Hotan}, A.~W. and {Lee-Waddell}, K. and {Madrid}, J.~P. and {Marvil}, J. and {McConnell}, D. and {Popping}, A. and {Voronkov}, M.~A. and {Whiting}, M.~T. and {Allen}, G.~R. and {Bock}, D.~C.-J. and {Brodrick}, D.~P. and {Cooray}, F. and {DeBoer}, D.~R. and {Diamond}, P.~J. and {Ekers}, R. and {Gough}, R.~G. and {Hampson}, G.~A. and {Harvey-Smith}, L. and {Hay}, S.~G. and {Hayman}, D.~B. and {Jackson}, C.~A. and {Johnston}, S. and {Koribalski}, B.~S. and {McClure-Griffiths}, N.~M. and {Mirtschin}, P. and {Ng}, A. and {Norris}, R.~P. and {Pearce}, S.~E. and {Phillips}, C.~J. and {Roxby}, D.~N. and {Troup}, E.~R. and {Westmeier}, T.},
        title = "{The Detection of an Extremely Bright Fast Radio Burst in a Phased Array Feed Survey}",
      journal = {\apjl},
         year = 2017,
        month = may,
       volume = {841},
       number = {1},
          eid = {L12},
        pages = {L12},
          doi = {10.3847/2041-8213/aa71ff},
archivePrefix = {arXiv},
       eprint = {1705.07581},
 primaryClass = {astro-ph.HE},
       adsurl = {https://ui.adsabs.harvard.edu/abs/2017ApJ...841L..12B}
}

@ARTICLE{CHIME2018,
       author = {{CHIME/FRB Collaboration} and {Amiri}, M. and {Bandura}, K. and {Berger}, P. and {Bhardwaj}, M. and {Boyce}, M.~M. and {Boyle}, P.~J. and {Brar}, C. and {Burhanpurkar}, M. and {Chawla}, P. and {Chowdhury}, J. and {Cliche}, J.-F. and {Cranmer}, M.~D. and {Cubranic}, D. and {Deng}, M. and {Denman}, N. and {Dobbs}, M. and {Fandino}, M. and {Fonseca}, E. and {Gaensler}, B.~M. and {Giri}, U. and {Gilbert}, A.~J. and {Good}, D.~C. and {Guliani}, S. and {Halpern}, M. and {Hinshaw}, G. and {H{\"o}fer}, C. and {Josephy}, A. and {Kaspi}, V.~M. and {Landecker}, T.~L. and {Lang}, D. and {Liao}, H. and {Masui}, K.~W. and {Mena-Parra}, J. and {Naidu}, A. and {Newburgh}, L.~B. and {Ng}, C. and {Patel}, C. and {Pen}, U.-L. and {Pinsonneault-Marotte}, T. and {Pleunis}, Z. and {Rafiei Ravandi}, M. and {Ransom}, S.~M. and {Renard}, A. and {Scholz}, P. and {Sigurdson}, K. and {Siegel}, S.~R. and {Smith}, K.~M. and {Stairs}, I.~H. and {Tendulkar}, S.~P. and {Vanderlinde}, K. and {Wiebe}, D.~V.},
        title = "{The CHIME Fast Radio Burst Project: System Overview}",
      journal = {\apj},
         year = 2018,
        month = aug,
       volume = {863},
       number = {1},
          eid = {48},
        pages = {48},
          doi = {10.3847/1538-4357/aad188},
archivePrefix = {arXiv},
       eprint = {1803.11235},
 primaryClass = {astro-ph.IM},
       adsurl = {https://ui.adsabs.harvard.edu/abs/2018ApJ...863...48C}
}

@ARTICLE{CHIMEFRB2023repeaters,
       author = {{Chime/Frb Collaboration} and {Andersen}, Bridget C. and {Bandura}, Kevin and {Bhardwaj}, Mohit and {Boyle}, P.~J. and {Brar}, Charanjot and {Cassanelli}, Tomas and {Chatterjee}, S. and {Chawla}, Pragya and {Cook}, Amanda M. and {Curtin}, Alice P. and {Dobbs}, Matt and {Dong}, Fengqiu Adam and {Faber}, Jakob T. and {Fandino}, Mateus and {Fonseca}, Emmanuel and {Gaensler}, B.~M. and {Giri}, Utkarsh and {Herrera-Martin}, Antonio and {Hill}, Alex S. and {Ibik}, Adaeze and {Josephy}, Alexander and {Kaczmarek}, Jane F. and {Kader}, Zarif and {Kaspi}, Victoria and {Landecker}, T.~L. and {Lanman}, Adam E. and {Lazda}, Mattias and {Leung}, Calvin and {Lin}, Hsiu-Hsien and {Masui}, Kiyoshi W. and {McKinven}, Ryan and {Mena-Parra}, Juan and {Meyers}, Bradley W. and {Michilli}, D. and {Ng}, Cherry and {Pandhi}, Ayush and {Pearlman}, Aaron B. and {Pen}, Ue-Li and {Petroff}, Emily and {Pleunis}, Ziggy and {Rafiei-Ravandi}, Masoud and {Rahman}, Mubdi and {Ransom}, Scott M. and {Renard}, Andre and {Sand}, Ketan R. and {Sanghavi}, Pranav and {Scholz}, Paul and {Shah}, Vishwangi and {Shin}, Kaitlyn and {Siegel}, Seth and {Smith}, Kendrick and {Stairs}, Ingrid and {Su}, Jianing and {Tendulkar}, Shriharsh P. and {Vanderlinde}, Keith and {Wang}, Haochen and {Wulf}, Dallas and {Zwaniga}, Andrew},
        title = "{CHIME/FRB Discovery of 25 Repeating Fast Radio Burst Sources}",
      journal = {\apj},
         year = 2023,
        month = apr,
       volume = {947},
       number = {2},
          eid = {83},
        pages = {83},
          doi = {10.3847/1538-4357/acc6c1},
archivePrefix = {arXiv},
       eprint = {2301.08762},
 primaryClass = {astro-ph.HE},
       adsurl = {https://ui.adsabs.harvard.edu/abs/2023ApJ...947...83C}
}

@ARTICLE{Cordes2019,
       author = {{Cordes}, James M. and {Chatterjee}, Shami},
        title = "{Fast Radio Bursts: An Extragalactic Enigma}",
      journal = {\araa},
         year = 2019,
        month = aug,
       volume = {57},
        pages = {417-465},
          doi = {10.1146/annurev-astro-091918-104501},
archivePrefix = {arXiv},
       eprint = {1906.05878},
 primaryClass = {astro-ph.HE},
       adsurl = {https://ui.adsabs.harvard.edu/abs/2019ARA&A..57..417C}
}

@ARTICLE{Petroff2016,
       author = {{Petroff}, E. and {Barr}, E.~D. and {Jameson}, A. and {Keane}, E.~F. and {Bailes}, M. and {Kramer}, M. and {Morello}, V. and {Tabbara}, D. and {van Straten}, W.},
        title = "{FRBCAT: The Fast Radio Burst Catalogue}",
      journal = {\pasa},
         year = 2016,
        month = sep,
       volume = {33},
          eid = {e045},
        pages = {e045},
          doi = {10.1017/pasa.2016.35},
archivePrefix = {arXiv},
       eprint = {1601.03547},
 primaryClass = {astro-ph.HE},
       adsurl = {https://ui.adsabs.harvard.edu/abs/2016PASA...33...45P}
}

@ARTICLE{2019MNRAS.489..919K,
       author = {{Kocz}, J. and {Ravi}, V. and {Catha}, M. and {D'Addario}, L. and {Hallinan}, G. and {Hobbs}, R. and {Kulkarni}, S. and {Shi}, J. and {Vedantham}, H. and {Weinreb}, S. and {Woody}, D.},
        title = "{DSA-10: a prototype array for localizing fast radio bursts}",
      journal = {\mnras},
         year = 2019,
        month = oct,
       volume = {489},
       number = {1},
        pages = {919-927},
          doi = {10.1093/mnras/stz2219},
archivePrefix = {arXiv},
       eprint = {1906.08699},
 primaryClass = {astro-ph.IM},
       adsurl = {https://ui.adsabs.harvard.edu/abs/2019MNRAS.489..919K}
}

@ARTICLE{Kirsten2020,
       author = {{Kirsten}, F. and {Snelders}, M.~P. and {Jenkins}, M. and {Nimmo}, K. and {van den Eijnden}, J. and {Hessels}, J.~W.~T. and {Gawro{\'n}ski}, M.~P. and {Yang}, J.},
        title = "{Detection of two bright radio bursts from magnetar SGR 1935 + 2154}",
      journal = {Nature Astronomy},
         year = 2021,
        month = apr,
       volume = {5},
        pages = {414-422},
          doi = {10.1038/s41550-020-01246-3},
archivePrefix = {arXiv},
       eprint = {2007.05101},
 primaryClass = {astro-ph.HE},
       adsurl = {https://ui.adsabs.harvard.edu/abs/2021NatAs...5..414K}
}

@ARTICLE{ZackayOfek2017,
       author = {{Zackay}, Barak and {Ofek}, Eran O.},
        title = "{An Accurate and Efficient Algorithm for Detection of Radio Bursts with an Unknown Dispersion Measure, for Single-dish Telescopes and Interferometers}",
      journal = {\apj},
         year = 2017,
        month = jan,
       volume = {835},
       number = {1},
          eid = {11},
        pages = {11},
          doi = {10.3847/1538-4357/835/1/11},
archivePrefix = {arXiv},
       eprint = {1411.5373},
 primaryClass = {astro-ph.IM},
       adsurl = {https://ui.adsabs.harvard.edu/abs/2017ApJ...835...11Z}
}

@ARTICLE{Lorimer2007,
       author = {{Lorimer}, D.~R. and {Bailes}, M. and {McLaughlin}, M.~A. and {Narkevic}, D.~J. and {Crawford}, F.},
        title = "{A Bright Millisecond Radio Burst of Extragalactic Origin}",
      journal = {Science},
         year = 2007,
        month = nov,
       volume = {318},
       number = {5851},
        pages = {777},
          doi = {10.1126/science.1147532},
archivePrefix = {arXiv},
       eprint = {0709.4301},
 primaryClass = {astro-ph},
       adsurl = {https://ui.adsabs.harvard.edu/abs/2007Sci...318..777L}
}

@ARTICLE{Spitler2016,
       author = {{Spitler}, L.~G. and {Scholz}, P. and {Hessels}, J.~W.~T. and {Bogdanov}, S. and {Brazier}, A. and {Camilo}, F. and {Chatterjee}, S. and {Cordes}, J.~M. and {Crawford}, F. and {Deneva}, J. and {Ferdman}, R.~D. and {Freire}, P.~C.~C. and {Kaspi}, V.~M. and {Lazarus}, P. and {Lynch}, R. and {Madsen}, E.~C. and {McLaughlin}, M.~A. and {Patel}, C. and {Ransom}, S.~M. and {Seymour}, A. and {Stairs}, I.~H. and {Stappers}, B.~W. and {van Leeuwen}, J. and {Zhu}, W.~W.},
        title = "{A repeating fast radio burst}",
      journal = {\nat},
         year = 2016,
        month = mar,
       volume = {531},
       number = {7593},
        pages = {202-205},
          doi = {10.1038/nature17168},
archivePrefix = {arXiv},
       eprint = {1603.00581},
 primaryClass = {astro-ph.HE},
       adsurl = {https://ui.adsabs.harvard.edu/abs/2016Natur.531..202S}
}

@ARTICLE{Bochenek2020,
       author = {{Bochenek}, C.~D. and {Ravi}, V. and {Belov}, K.~V. and {Hallinan}, G. and {Kocz}, J. and {Kulkarni}, S.~R. and {McKenna}, D.~L.},
        title = "{A fast radio burst associated with a Galactic magnetar}",
      journal = {\nat},
         year = 2020,
        month = nov,
       volume = {587},
       number = {7832},
        pages = {59-62},
          doi = {10.1038/s41586-020-2872-x},
archivePrefix = {arXiv},
       eprint = {2005.10828},
 primaryClass = {astro-ph.HE},
       adsurl = {https://ui.adsabs.harvard.edu/abs/2020Natur.587...59B}
}

%%%%%%%%%%%%%%%%%%%%%%%%%%%%%%%%%%%%%%%%%%%%%%%%%%

%%%%%%%%%%%%%%%%% APPENDICES %%%%%%%%%%%%%%%%%%%%%

\appendix

\section{Candidate LIMBO FRBs} \label{appx:candidates}

\begin{figure*}
    \centering
    \includegraphics[width=0.9\linewidth]{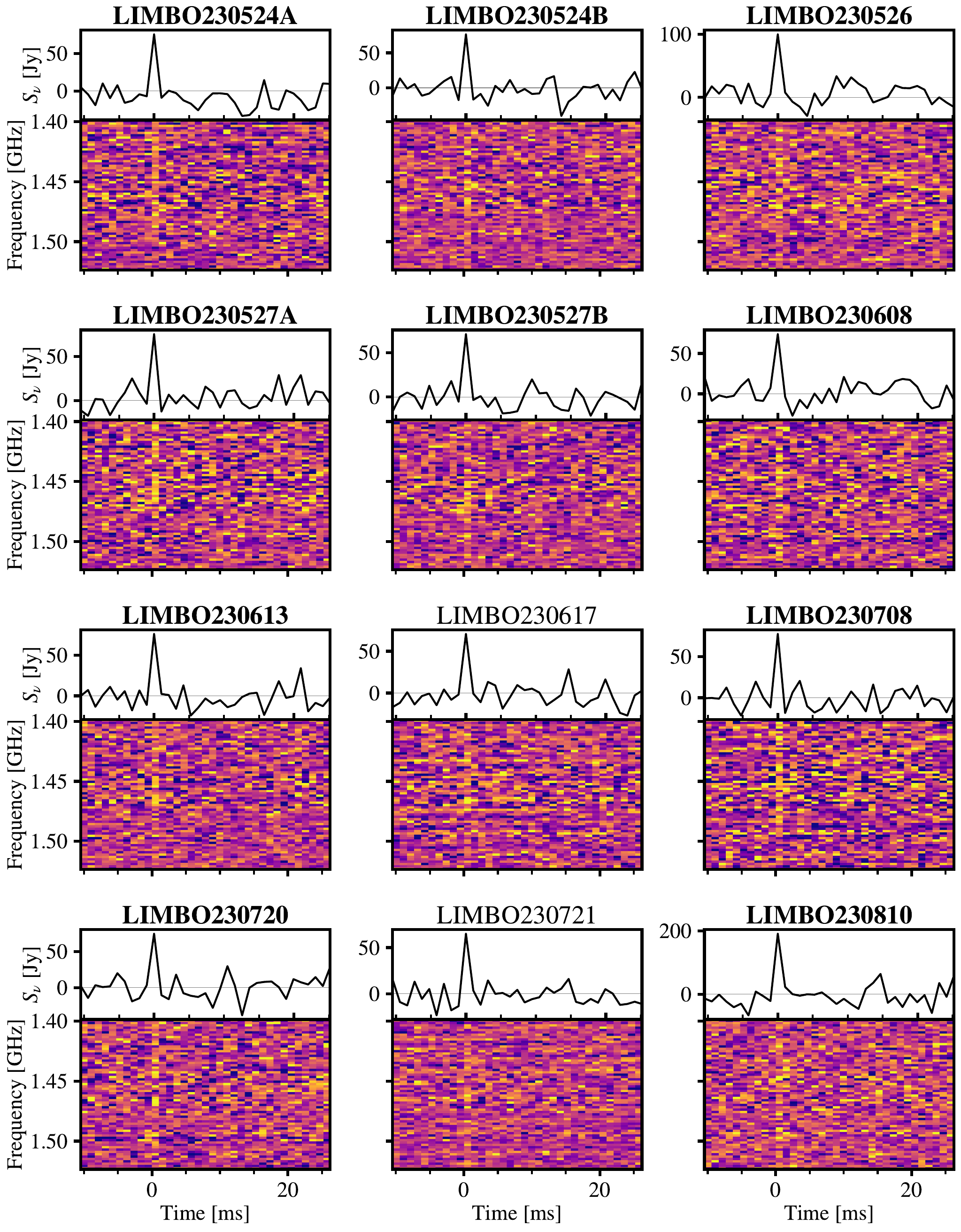}
    \caption{The power spectra of bursts detected by LIMBO from Galactic magnetar \sgr, tentatively classified as candidate FRBs. In each figure, the top panel shows the frequency-averaged flux density as a function of time, while the bottom panel displays the burst in both frequency and time domain. Bursts with bolded names represent the events we are most confident in classifying as candidate FRB detections. Each pulse is dedispersed to maximize its SNR. Pulse properties are given in Table \ref{tab:frb_props}.}
    \label{fig:hits}
\end{figure*}

Of the 24 events with detection significances $\geq 5.6$ obtained during the May-August 2023 LIMBO observing campaign of \sgr, 12 events are classified as candidate FRB detections. Their power spectra are shown in Figure \ref{fig:hits}, where each pulse is dedispersed to maximize its SNR. Pulse properties, such as dispersion measure, fluence, SNR, and Z-score are provided in Table \ref{tab:frb_props}.

\section{RFI-Induced False-Positive Triggers} \label{appx:rfis}

\begin{figure*}
    \centering
    \includegraphics[width=0.9\linewidth]{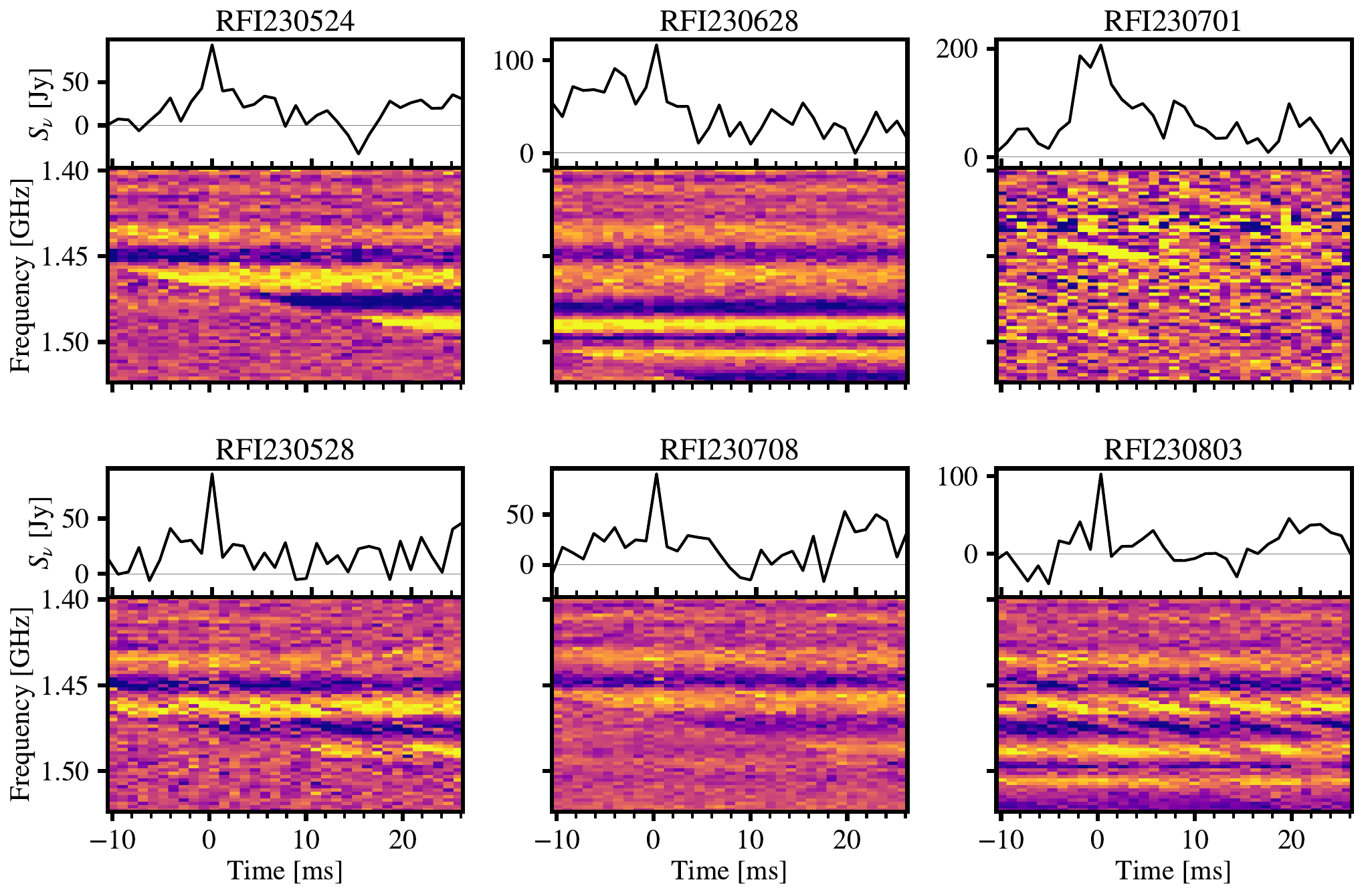}
    \caption{\emph{Top:} Example spectra from typical RFI-induced false-positive triggers from the LIMBO system. The diagonal contrast seen in the dynamic spectrum of RFI230524 and RF230628 is an artifact of the LIMBO RFI excision pipeline's flagging and inpainting procedures. The variability in both time and frequency are quite different from the behaviour expected of FRBs, making it easy to visually distinguish these events as false-positive triggers. \emph{Bottom:} RFI-classified events with possible underlying FRB candidates. Inpainting and RFI contamination make analysing these events difficult. These events are therefore excluded from the larger FRB candidate sample as LIMBO does not yet have means to recover FRB-like bursts from files heavily contaminated with overlying RFI.}
    \label{fig:misses}
\end{figure*}

Of the remaining 12 events with $Z\geq5.6$, 7 are identified to be consistent with false-positive triggers due to spurious RFI contamination. Three representative spectra are shown in the top row of Figure \ref{fig:misses}. Broad or variable time and frequency structure exhibited in these spectra is inconsistent with that expected for FRBs, motivating their removal from the detection sample. Diagonal, high-contrast features in the dynamic spectra are artifacts of the LIMBO RFI-rejection pipeline, which flags RFI-contaminated regions, inpaints them with Gaussian noise, and subsequently dedisperses the data.

The remaining 3 candidates exhibit mixed morphology, combining FRB-like features with clear signatures of RFI (Figure \ref{fig:misses} (\emph{bottom})). Because these events are more ambiguous and LIMBO does not currently implement a recovery pipeline for these types of events, they are also removed from the final detection sample.

%%%%%%%%%%%%%%%%%%%%%%%%%%%%%%%%%%%%%%%%%%%%%%%%%%

% Don't change these lines
\bsp	% typesetting comment
\label{lastpage}
\end{document}